\documentclass[fleqn,usenatbib]{mnras}

\usepackage{newtxtext,newtxmath}
\usepackage{mathtools}

\usepackage[T1]{fontenc}

\DeclareRobustCommand{\VAN}[3]{#2}
\let\VANthebibliography\thebibliography
\def\thebibliography{\DeclareRobustCommand{\VAN}[3]{##3}\VANthebibliography}

\usepackage{graphicx}	
\usepackage{amsmath}	
\usepackage{xspace}
\usepackage{ulem}
\makeatletter 
  \patchcmd{\NAT@citex}
    {\@citea\NAT@hyper@{%
      \NAT@nmfmt{\NAT@nm}%
      \hyper@natlinkbreak{\NAT@aysep\NAT@spacechar}{\@citeb\@extra@b@citeb}%
      \NAT@date}}
    {\@citea\NAT@nmfmt{\NAT@nm}%
    \NAT@aysep\NAT@spacechar\NAT@hyper@{\NAT@date}}{}{}

  \patchcmd{\NAT@citex}
    {\@citea\NAT@hyper@{%
      \NAT@nmfmt{\NAT@nm}%
      \hyper@natlinkbreak{\NAT@spacechar\NAT@@open\if*#1*\else#1\NAT@spacechar\fi}%
        {\@citeb\@extra@b@citeb}%
      \NAT@date}}
    {\@citea\NAT@nmfmt{\NAT@nm}%
    \NAT@spacechar\NAT@@open\if*#1*\else#1\NAT@spacechar\fi\NAT@hyper@{\NAT@date}}
    {}{}
\makeatother

\newcommand{\hMsol}{h^{-1}\,{\rm M_\odot}}
\newcommand{\Msun}{{\rm M_\odot}}

\newcommand{\hMpc}{h^{-1}\,{\rm Mpc}}

\newcommand{\ckpc}{{\rm ckpc}}
\newcommand{\cMpc}{{\rm cMpc}}
\newcommand{\kpc}{{\rm kpc}}

\newcommand{\Mh}{M_{\rm h}}

\newcommand{\Lbox}{L_\mathrm{box}}

\newcommand{\tzero}{$T_0$\xspace}
\newcommand{\arepo}{{\sc arepo}\xspace}
\newcommand{\areport}{{\sc arepo-rt}\xspace}
\newcommand{\eg}{e.g.\xspace}
\newcommand{\ie}{i.e.\xspace}
\newcommand{\hii}{\ion{H}{II}\xspace}
\newcommand{\thesan}{\textsc{thesan}\xspace}
\newcommand{\jwst}{\texttt{JWST}\xspace}
\newcommand{\alma}{\texttt{ALMA}\xspace}
\newcommand{\hst}{\texttt{HST}\xspace}
\newcommand{\ska}{\texttt{SKA}\xspace}
\newcommand{\hera}{\texttt{HERA}\xspace}
\newcommand{\lofar}{\texttt{LOFAR}\xspace}
\newcommand{\thesanone}{\mbox{\textsc{thesan-1}}\xspace}
\newcommand{\thesantwo}{\mbox{\textsc{thesan-2}}\xspace}
\newcommand{\thesanwc}{\mbox{\textsc{thesan-wc-2}}\xspace}
\newcommand{\thesanlow}{\mbox{\textsc{thesan-low-2}}\xspace}
\newcommand{\thesanhigh}{\mbox{\textsc{thesan-high-2}}\xspace}

\newcommand{\thesansdao}{\mbox{\textsc{thesan-sdao-2}}\xspace}
\newcommand{\thesantng}{\mbox{\textsc{thesan-tng-2}}\xspace}
\newcommand{\thesantngsdao}{\mbox{\textsc{thesan-tng-sdao-2}}\xspace}
\newcommand{\thesannort}{\mbox{\textsc{thesan-nort-2}}\xspace}
\newcommand{\thesandarkone}{\mbox{\textsc{thesan-dark-1}}\xspace}
\newcommand{\thesandarktwo}{\mbox{\textsc{thesan-dark-2}}\xspace}
\newcommand{\lyalpha}{$\mathrm{Ly}\alpha$\xspace}
\newcommand{\dd}{\mathrm{d}}

\title[Introducing the {\sc thesan} project]{Introducing the \thesan project: radiation-magnetohydrodynamic simulations of the epoch of reionization}

\author[R.\ Kannan et al.]{%
R.~Kannan,$^{1}$\thanks{E-mail: \href{mailto:rahul.kannan@cfa.harvard.edu}{rahul.kannan@cfa.harvard.edu}}
E.~Garaldi,$^{2}$\thanks{E-mail: \href{mailto:egaraldi@mpa-garching.mpg.de}{egaraldi@mpa-garching.mpg.de}}
A.~Smith,$^{3}$\thanks{E-mail: \href{mailto:arsmith@mit.edu}{arsmith@mit.edu}; NHFP Einstein Fellow.}
R.~Pakmor,$^{2}$
V.~Springel,$^{2}$
M.~Vogelsberger$^{3}$
and L.~Hernquist$^{1}$
\\
$^{1}$Center for Astrophysics $\vert$ Harvard $\&$ Smithsonian, 60 Garden Street, Cambridge, MA 02138, USA \\
$^{2}$Max-Planck Institute for Astrophysics, Karl-Schwarzschild-Str.~1, D-85741 Garching, Germany \\
$^{3}$Department of Physics, Massachusetts Institute of Technology, Cambridge, MA 02139, USA 
}

\date{Accepted XXX. Received YYY; in original form ZZZ}

\pubyear{2021}

\begin{document}
\label{firstpage}
\pagerange{\pageref{firstpage}--\pageref{lastpage}}
\maketitle

\begin{abstract}
We introduce the \thesan project, a suite of large volume ($L_\mathrm{box} = 95.5 \, \mathrm{cMpc}$) radiation-magneto-hydrodynamic simulations that simultaneously model the large-scale statistical properties of the intergalactic medium (IGM) during reionization and the resolved characteristics of the galaxies responsible for it. The flagship simulation has dark matter and baryonic mass resolutions of $3.1 \times 10^6\,\Msun$ and $5.8 \times 10^5\,\Msun$, respectively. The gravitational forces are softened on scales
of $2.2$~ckpc with the smallest cell sizes reaching $10$\,pc at $z=5.5$, enabling predictions down to the atomic cooling limit. The simulations use an efficient radiation hydrodynamics solver (\areport) that precisely captures the interaction between ionizing photons and gas, coupled to well-tested galaxy formation (IllustrisTNG) and dust models to accurately predict the properties of galaxies. Through a complementary set of medium resolution simulations we investigate the changes to reionization introduced by different assumptions for ionizing escape fractions, varying dark matter models, and numerical convergence. The fiducial simulation and model variations are calibrated to produce realistic reionization histories that match the observed evolution of the global neutral hydrogen fraction and electron scattering optical depth to reionization. They also match a wealth of high-redshift observationally inferred data, including  the stellar-to-halo-mass relation, galaxy stellar mass function, star formation rate density, and the mass-metallicity relation, despite the galaxy formation model being mainly calibrated at $z = 0$. We demonstrate that different reionization models give rise to varied bubble size distributions that imprint unique signatures on the 21\,cm emission, especially on the slope of the power spectrum at large spatial scales, enabling current and upcoming 21\,cm experiments to accurately characterise the sources that dominate the ionizing photon budget.
\end{abstract}

\begin{keywords}
galaxies: high-redshift -- cosmology: dark ages, reionization, first stars -- radiative transfer -- methods: numerical
\end{keywords}



\section{Introduction}

In the aftermath of
the Big Bang, the Universe was comprised of a hot dense plasma of matter
and radiation. As the gas expanded and cooled, the protons began to capture the free electrons and form atomic hydrogen. These recombinations diminished the number density of free electrons allowing the matter and radiation to decouple, making the Universe transparent to light. The Universe then went through a period of darkness with no new sources of visible light.
After hundreds of millions of years gravitational forces had amplified small density fluctuations imprinted during the Big Bang to initiate the formation of stars and proto-galaxies. These first stars and galaxies had an important effect on the surrounding environment. They emitted copious amounts of Lyman continuum (LyC; $> 13.6$\,eV) radiation, ionizing and heating up the neutral intergalactic medium (IGM) in-between galaxies \citep[e.g. ][]{Shapiro1987, Haardt2012, Gnedin2000}. This phase transition known as the Epoch of Reionization (EoR) represents an important evolutionary link between the smooth matter distribution at early times as revealed by the cosmic microwave background (CMB) and the large-scale structure observed today. Several important questions about this epoch remain unanswered, including, what are the properties of the main sources that reionized the Universe, how long was the process, when did it come to an end, what temperature was the IGM heated to, and how effective was this heating in regulating star formation in low mass galaxies?

Gaining insights into the EoR has been quite challenging on both theoretical and observational fronts. The galaxy population at very high redshifts
is largely unexplored, because even the deepest Hubble Space Telescope (\hst) observations have been able to detect only about a thousand galaxy candidates between $z = 6$--$8$ and only a few at higher redshifts \citep[e.g.][]{Bouwens2015, Livermore2017, Atek2018}. Most of these galaxies do not have a spectroscopic confirmation/characterization and only a handful of objects have been observed with complementary facilities such as the Atacama Large millimetre Array \citep[\alma; ][]{Decarli2018, Hashimoto2018}. Hence, current observational constraints provide only weak tests of reionization-era galaxy-formation models.  Studying the ionization and temperature structure of the intergalactic medium (IGM) during reionization is even more challenging because most of it is at a very low density. Current instruments do not have the sensitivity to detect the IGM in emission. Using bright sources (such as high-redshift quasars) as backlights, the state of the IGM can, however, be studied in absorption through the Lyman-alpha (\lyalpha) transition of the hydrogen atom \citep{Becker2001, Fan2006, Becker2015, Davies2018}. Historically, this technique has been limited to the post-reionization Universe as a consequence of the very-large cross-section of the \lyalpha transition. However, in recent years, improvements in instrumentation have allowed us to push these kind of observations deep into the tail-end of reionization, unveiling features that are largely unexplored \citep[\eg][]{Becker2015,Barnett+2017,Chardin+2018,Eilers+2018,Bosman+2018,Garaldi+2019croc,Yang+2020}. Unfortunately, it is still difficult to build a comprehensive picture of the state of the IGM based on the information from individual quasar sightlines alone.

However, the state of the field is about to change due to a suite of instruments and telescopes that will become available in the next few years. For example, the James Webb Space Telescope (\jwst) with its large mirror and infrared frequency coverage, will enable the discovery of fainter and higher-redshift galaxies (up to $z\sim 20$) and constrain their rest-UV/optical emission \citep{Kalirai2018, Williams2018}. Additionally, the fragmented view of the IGM will soon be replaced by a more complete picture of the neutral hydrogen distribution in the Universe by current and upcoming radio telescopes including the Low-Frequency Array (\lofar), Square Kilometer Array (\ska) and Hydrogen Epoch of Reionization Array (\hera). This will be achieved using the line intensity mapping (LIM) technique, which will measure the spatial fluctuations in the integrated emission of the 21\,cm spin flip transition of the hydrogen atom \citep{Mellema2006, PAPER, LOFAR, HERA}. In a similar vein, there are plans to use LIM of spectral lines originating from many individually unresolved galaxies to faithfully trace the growth and evolution of large-scale structures (e.g. Cerro Chajnantor Atacama Telescope-prime: \texttt{CCAT-p}; Spectro-Photometer for the history of the Universe, Epoch of Reionization and Ices Explorer: \texttt{SPHEREx}). While traditional galaxy surveys probe discrete objects, whose emission is bright enough to be imaged directly, LIM is sensitive to all sources of emission in the line and thus enables the universal study of galaxy formation and evolution \citep[see][for an overview of current and upcoming experiments]{Kovetz2019}. These advancements will unleash a flood of high-$z$ observational data that promises to usher in a new era of cosmic reionization studies.

It is clear that the Astronomy community is devoting significant resources in the coming decade(s) to study high-redshift structure formation and reionization. It is, therefore, imperative that theoretical/numerical models achieve sufficient accuracy and physical fidelity to meaningfully interpret these new observational results \citep{Dayal2018}. Modelling reionization is especially challenging because it is a uniquely multiphysics and multiscale process. The formation of the first stars and black holes and the subsequent production of radiation
occurs at
sub-parsec scales. The escape of radiation from these sources and the corresponding propagation out of galaxies is modulated by the physics of structure formation on parsec to kpc scales. Finally, the
interaction between these photons and the low-density IGM
transforms the entire Universe from a cold-neutral state to a hot-ionized one. Therefore, gaining insights into this process requires models that are able to make self-consistent predictions over a dynamic range of $\sim 10^{10}$. Moreover, a comprehensive characterisation of this processes requires an accurate description of a plethora of physical processes relevant for the formation of galaxies, such as the
behaviour
of dark matter, the dynamics of gas flows, the physics of star and black hole formation and feedback, as well as radiation hydrodynamics. It is, therefore, virtually impossible to treat this process analytically making numerical radiation hydrodynamic (RHD) simulations an essential tool to investigate the EoR.

The computational cost of fully coupled RHD simulations has forced many works to use a combination of linear perturbation theory to evolve the matter density field and an excursion-set formalism to  model the \hii regions \citep[see for e.g.][]{Furlanetto2004, 21cmfast, Fialkov2020, Munoz2020}. More accurate (but still computationally efficient) representations of the reionization process are obtained by populating dark matter halos (from pure N-body simulations) with either
idealized
models of galaxy formation and post-processing them with radiative transfer calculations \citep[RT;][]{Ciardi2003, Iliev2007, McQuinn2007, McQuinn2009} or using semi-analytic galaxy evolution models that incorporate the spatial reionization process on the fly \citep{Mutch2016, Seiler2019, Hutter2021}. These methods provide a  viable pathway to model the large representative volumes ($\sim 500$ cMpc) and efficiently explore a wider 
parameter space. Unfortunately, they require quite a few tuneable free parameters such as the star formation history, source functions, escape fractions, and most importantly, the gas density distribution, including assumptions about small-scale gas clumping factors. 

More recent efforts have
advanced this approach to include
post-processing hydrodynamics+N-body simulations with RT to model reionization \citep{Ciardi2012, Bauer2015, Kulkarni2017, Kulkarni2019, Eide2020, Keating2020}. They employ the correct gas density distribution and, if the star formation and feedback model is well constrained, a realistic source function. However, they are still unable to capture the small-scale coupling between the photons and the gas, leading to degeneracies that reduce the fidelity of the predictions. Simulations that include fully coupled RHD are numerically challenging, but allow us to model reionization from first principles. They have been quite successful in shedding light on the sources of reionization \citep[e.g. the role of binary stars; ][]{Rosdahl2018}, as well as accurately capturing the role of photoheating feedback during reionization on low mass galaxies \citep[$\lesssim 10^9 \, \mathrm{M}_\odot$; ][]{Gnedin2014, Pawlik2017} and the state of the ISM of high-$z$ galaxies \citep{Pallottini2017, Katz2019}. 

Unfortunately, these models are unable to paint a complete picture of the reionization process. Large volume simulations ($\gtrsim 100$\,cMpc) are the ideal tool for studying the statistical properties of the reionization observables but lack the necessary resolution to accurately model the sources responsible for it \citep{Pawlik2017, Ocvirk2020}. On the other hand, high-resolution small volume/zoom-in
simulations are well-suited for understanding the small-scale properties of the sources \citep{Xu2016, Pallottini2017} but they generally lack the ability to translate results to larger scales. More importantly, the galaxy formation models used in these simulations are not sufficiently well tested  \citep{Gnedin2014, Xu2016, Ceverino2017, Rosdahl2018, Trebitsch2020a} because they have been specifically designed and evaluated in simulations that mainly predict observables above $z \gtrsim 6$, where the constraints are fairly limited. When some of these galaxy formation models are used to simulate galaxies down to lower redshifts, they give rise to galaxy populations that are incompatible with the observed ones. For example, they produce an order of magnitude more stars than what is expected \citep{Trebitsch2020b} and bulge dominated galaxies with centrally peaked rotation curves \citep{Mitchell2021}. 

In this, and two initial companion papers \citep{GaraldiThesan, SmithThesan}, we present a novel simulation campaign designed to efficiently leverage current and upcoming high-redshift observations to constrain the physics of reionization. We use \arepo \citep{Springel2010} to simulate a relatively large representative volume of the Universe ($\mathrm{L_{box}} \sim 95.5$\,cMpc) with enough mass and spatial resolution to properly model atomic cooling halos ($\mathrm{M_{halo} \sim 10^8\,M_\odot/}h$) throughout the entire simulation volume. We use state-of-the-art and well-tested galaxy formation (IllustrisTNG; \citealt{Springel2018, Marinacci2018, Naiman2018, Nelson2018, Pillepich2018}) and dust models \citep{McKinnon2017} to accurately predict the properties of the sources (such as stars, galaxies and blackholes) responsible for the reionization process. This is coupled to a novel efficient and accurate radiation hydrodynamics and non-equilibrium thermochemistry solver (\areport; \citealt{Kannan2019}) to track the evolution of the ionization fronts and their back-reaction on galaxy formation. Additionally, we include a suite of medium resolution simulations intended to investigate the changes to reionization induced by different LyC escape fractions from galaxies, varying dark matter models, the impact of assuming a constant radiation background in galaxy formation simulations and numerical convergence. The methodology is introduced in Section~\ref{sec:methods} with the main results presented in Section~\ref{sec:results} and the conclusions outlined in Section~\ref{sec:conclusions}.

\section{Methods}
\label{sec:methods}

Simulations presented in this work are performed using {\sc arepo-rt} \citep{Kannan2019}, a novel radiation hydrodynamic extension to the moving mesh hydrodynamics code {\sc arepo} \citep{Springel2010, Arepo-public}\footnote{Public code access and documentation available at \href{https://arepo-code.org}{\texttt{www.arepo-code.org}}.}. It solves the (radiation-magneto-) hydrodynamical equations on an unstructured mesh, built from the Voronoi tessellation of a set of mesh-generating points which follow the flow of gas. A quasi-Lagrangian solution to the hydrodynamic equations is achieved by solving them at interfaces between moving mesh cells in the rest frame of the interface. Higher-order accuracy is achieved by using second order Runge-Kutta time integration coupled with a least square fit (LSF) gradient estimate that performs well even on highly distorted meshes \citep{Pakmor2016}. Additionally, the mesh is frequently regularised according to the algorithm described in \citet{Vogelsberger2012}. The magnetic fields are evolved using the ideal MHD equations outlined in \citet{Pakmor2013} and an eight-wave formalism is used to control divergence errors \citep{Powell1999}. The semi-Lagrangian character of {\sc arepo} allows it to naturally adjust the mesh resolution to the underlying density field and is therefore well suited to simulate systems with a large dynamical range.

Gravity is solved using the Hybrid Tree-PM approach which estimates the short range forces using a hierarchical oct-tree algorithm \citep{Barnes1986}, while the long range forces are computed using the particle mesh method in which the gravitational potential is obtained by binning particles into a grid of density values and then solving the Poisson equation using the Fourier method. Additionally, if the number of active particles is below a certain threshold, then the gravitational force is calculated by a direct summation. This is a particularly attractive option in large scale simulations, where the computational cost of direct summation is lower than the tree algorithm for the lowest timebins, which are generally populated by just a few active particles. A hierarchical time integration approach, which allows us to construct the tree only for the currently active particle set is used to speed up the gravity calculations, especially since the timebin hierarchy can get very deep in our highest resolution simulations \citep{Gadget4}. Finally, we note that in order to overcome the correlated force errors at large node boundaries arising from the nearly static particle distributions at high redshift in cosmological simulations, we employ the simple yet effective method of randomising the relative location of the particle set with respect to the computational box each time a new domain decomposition is computed \citep{Gadget4}.

\subsection{The radiation hydrodynamics solver}
The propagation of radiation is handled using a moment-based approach to solve the radiation transport equations \citep{Kannan2019}. We solve the set of coupled hyperbolic conservation equations for photon number density and photon flux. This set of equations is closed using the M1 scheme \citep{Levermore1984, Dubroca1999}. The Riemann problem is solved at each cell interface by computing the flux using Godunov’s approach \citep{Godunov1959}. We achieve second order accuracy by replacing the piecewise constant (PC) approximation of Godunov's scheme with a slope-limited piecewise linear spatial extrapolation and a half timestep, first order time extrapolation, obtaining the primitive variables on both sides of the interface \citep{vanLeer1979}. We perform the spatial extrapolations using a local LSF gradient estimate \citep{Pakmor2016}. 

We divide the UV continuum into three frequency bins relevant for hydrogen and helium photoionization
grouping radiation between energy intervals of $[13.6, 24.6, 54.4,\infty)$\,eV.
For each frequency bin `$i$', we evolve the comoving photon number density ($\tilde{N}_i$) and photon flux (${\bf{\tilde{F}}}_i$), which are defined as $\tilde{N}_i = a^3 N_i$ and ${\bf{\tilde{F}}}_i = a^3 {\bf{F}}_i$, where $N_i$ and ${\bf{F}}_i$ are the physical photon number density and flux, respectively. The cube of the scale factor ($a$) is multiplied to the physical quantities to account for the loss of photon energy due to cosmological expansion. Assuming that the Universe does not expand significantly before a photon is absorbed \citep{Gnedin2001}, the transport equations take the form \citep[see also][]{Wu2019a}
\begin{equation}
     \frac{\partial \tilde{N}_i}{\partial t} + \frac{1}{a} \nabla \cdot \tilde{\bf{F}}_i = 0 \, ,
     \end{equation}
     \begin{equation}
     \frac{\partial {{\tilde{\bf{F}}}}_i} {\partial t} + \frac{\tilde{c}^2}{a} \nabla \cdot {\tilde{\mathbb{P}}_i} = 0 \,, \\
\end{equation}
where, $\tilde{c}$ is the reduced speed of light, which for our runs is set to $\tilde{c} = 0.2\,c$ (where $c$ is the speed of light in vacuum),
and $\tilde{\mathbb{P}}_i$ is the comoving pressure tensor, which is related to the photon number density by the Eddington tensor. In Appendix~\ref{app:RSLA} we include a discussion on the reduced speed of light approximation in the context of \thesan.
 
 \begin{table*}
	\centering
	\caption{Table outlining the frequency discretisation of the radiation field as used in our simulations. It lists the frequency range (first column), the mean photoionization cross section ($\sigma$) for the different species ($\ion{H}{I}$, second column; $\ion{He}{I}$, third column; $\ion{He}{II}$, fourth column), the energy injected into the gas per interacting photon ($\mathcal{E}$) for the different species ($\ion{H}{I}$, fifth column; $\ion{He}{I}$, sixth column; $\ion{He}{II}$, seventh column) and the mean energy per photon ($e$; eighth column).}
	\label{table:radiation}
	\begin{tabular}{ccccccccccc} 
		\hline
		Frequency bin & $\sigma_\ion{H}{I}$ & $\sigma_\ion{He}{I}$ & $\sigma_\ion{He}{II}$ & $\mathcal{E}_\ion{H}{I}$ & $\mathcal{E}_\ion{He}{I}$  & $\mathcal{E}_\ion{He}{II}$ & $e$\\  
		  $[$eV$]$& [cm$^2$] & [cm$^2$] & [cm$^2$]  & [eV] & [eV] & [eV] & [eV]\\
		\hline
		$13.6 - 24.6$ & $3.31 \times 10^{-18}$ & 0 & 0 & 3.25 & 0 & 0 & 18.17\\
		$24.6 - 54.4$ & $6.99 \times 10^{-19}$ & $4.53 \times 10^{-18}$ & 0 & 15.64 & 4.74 & 0 & 32.05\\
		$54.4 - \infty$ & $1.09 \times 10^{-19}$ & $7.73 \times 10^{-19}$  & $1.42 \times 10^{-18}$ & 42.86 & 31.87 & 2.10 & 56.99\\
		\hline
	\end{tabular}
\end{table*}
The radiation fields are coupled to the gas via a non-equilibrium thermochemistry module. It incorporates chemistry and cooling by atomic hydrogen and helium ($\Lambda_p$), equilibrium cooling from metals ($\Lambda_M$) and Compton cooling ($\Lambda_C$),
\begin{equation}
 \Lambda  = \Lambda_p(n_j, N_i, T) + \frac{Z}{Z_\odot} \Lambda_M(\rho, T, z) + \Lambda_C(\rho, T,z) \,,
\end{equation}
where $n_j \in [n_\ion{H}{I}, n_\ion{H}{II}, n_\ion{He}{I}, n_\ion{He}{II}, n_\ion{He}{III}]$ are the number densities of the ionic species tracked by the thermochemistry module, $\rho$ is the density of the gas, $Z$ is the metallicity (with $Z_\odot$ representing the metallicity of the Sun), $T$ is the temperature and $z$ is the redshift. The primordial chemistry and cooling are solved according to the non-equilibrium thermochemistry module described in Section 3.2.1 of \citet{Kannan2019}, metal cooling is implemented assuming ionization equilibrium for a given portion of dust-free and optically thin gas in a UV background radiation field given by \citet{CAFG2009}. Practically, this cooling rate is computed from a look-up table containing the pre-calculated cooling values computed from CLOUDY (see \citealt{Vogelsberger2013}, for more details). We note that the metal cooling rate assumes a certain metagalactic UV background, which is spatially uniform. In our current reionization simulations, this approximation is certainly not valid. However, we have checked that this approximation has a negligible affect on our results. We note that the tracked ionic species ($n_j$) are advected along with the gas motions, so as to preserve the ionization state of the gas from one timestep to another. This is necessary in order to preserve the true non-equilibrium nature of our ionization solver.

Both stars and active galactic nuclei (AGN) are considered sources of radiation. The luminosity and spectral energy density of stars is a complex function of age and metallicity taken from the Binary Population and Spectral Synthesis models (BPASS; \citealt{BPASS2017}). The radiation from AGN on the other hand is scaled linearly with the mass accretion rate with a radiative conversion efficiency of 0.2 \citep{Weinberger2018}. The AGN spectral energy distribution (SED) uses the \citet{Lusso2015} parametrization with $35.5$ per cent of the bolometric AGN luminosity at energies above $13.6$\,eV. The mean photoionization cross section ($\sigma$), the energy injected into the gas per interacting photon  ($\mathcal{E}$), and mean photon energy ($e$) vary from cell to cell due to the differing shapes of the radiation spectrum from different age and metallicity sources. The relatively low frequency resolution used in this work is not able to track this change in shape. We note that for the BPASS spectra the calculated radiation parameters are roughly constant and do not vary significantly with the metallicity and age of the star \citep{Rosdahl2018}. Moreover, we expect the contribution from black-holes to the ionizing photon budget to be minimal \citep{Parsa2018}. We therefore calculate them using a $2$\,Myr spectrum at quarter solar metallicity and employ the same values for all cells throughout the simulation as listed in Table~\ref{table:radiation}.

\subsection{The IllustrisTNG galaxy formation model}
Galaxies are predicted by many works to be the main source of ionizing photons in the high-redshift ($z \gtrsim 4$) Universe \citep{Haardt2012}, and therefore any simulation hoping to self-consistently reproduce the cosmic reionization history is required to properly model their formation and evolution. While the dynamics of dark matter, gas, stars and photons are all self-consistently simulated using \areport, processes occurring on scales smaller than the resolution limit of the simulation can not be captured self consistently. For this reason, many of the small scale processes (e.g.~star formation, black hole accretion, etc.) need to be included as empirical prescriptions. In our simulations, we employ the state-of-the-art IllustrisTNG galaxy formation model, which updates the previous Illustris simulations \citep{Vogelsberger2014} with a revised kinetic AGN feedback model for the low accretion state \citep{Weinberger2017} and an improved parameterisation of galactic winds \citep{Pillepich2018}. It includes: (i) a sub-resolution treatment of the interstellar medium (ISM) as a two-phase gas where cold clumps are embedded in a smooth, hot phase produced by supernova explosions \citep{Springel2003}; (ii) feedback from supernova explosions and stellar winds, in the form of kinetic and thermal energy; (iii) the production and evolution of nine elements (H, He, C, N, O, Ne, Mg, Si and Fe), as well as the tracking of the overall gas metallicity; (iv) density-, redshift-, metallicity- and temperature-dependent cooling; and (v) black-hole formation (via a seeding prescription), growth and feedback in two different regimes (quasar- and radio-mode). The model has been extensively tested in large-scale simulations, and is able to produce realistic galaxies that match, among others, the observed galactic color bimodality \citep{Nelson2018}, color-dependent spatial distribution and clustering of galaxies \citep{Springel2018}, galaxy mass function, galaxy sizes \citep{Genel2018, Pillepich2018b}, metal distributions \citep{Naiman2018, Vogelsberger2018}, magnetic field strength and structure \citep{Marinacci2018} and galaxy morphologies \citep{RG19, Tacchella2019}. All these elements are key to realistically simulate the reionization history of the Universe, that heavily depends on the source properties, and to investigate the correlation between small and large scales in the Universe.

\begin{table*}
	\centering
	\caption{\thesan simulation suite: From left to right the columns indicate the name of the simulation, boxsize, initial particle number, mass of the dark matter and gas particles, the (minimum) softening length of (gas) star and dark matter particles, minimum cell size at $z=5.5$, the final redshift, the escape fraction of ionizing photons from the birth cloud (if applicable) and a short description of the simulation.  }
	\label{table:simulations}
	\begin{tabular}{lccccccccc} 
		\hline
		Name & $L_\mathrm{box}$ & $N_\mathrm{particles}$ & $m_\mathrm{DM}$ & $m_\mathrm{gas}$ & $\epsilon$ & $r^\mathrm{min}_\mathrm{cell}$& $z_\mathrm{end}$ & $f_\mathrm{esc}$ & Description\\  
		& [cMpc] & & [$\mathrm{M}_\odot$] & [$\mathrm{M}_\odot$] & [ckpc] & [pc] & & &\\
		\hline
		\thesanone & $95.5$  & $2 \times 2100^3$ & $3.12 \times 10^6$ & $5.82 \times 10^5$ & $2.2$ & $\sim 10$ & $5.5$ & $0.37$ & fiducial \\
		\\
		\thesantwo & $95.5$  & $2 \times 1050^3$ & $2.49 \times 10^7$ & $4.66 \times 10^6$ & $4.1$ & $\sim 35$ & $5.5$ & $0.37$ & fiducial\\
		\thesanwc & $95.5$  & $2 \times 1050^3$ & $2.49 \times 10^7$ & $4.66 \times 10^6$ & $4.1$ & $\sim 35$ & $5.5$ & $0.43$ & weak convergence of $x_\mathrm{HI} (z)$ \\
		\thesanhigh & $95.5$  & $2 \times 1050^3$ & $2.49 \times 10^7$ & $4.66 \times 10^6$ & $4.1$ & $\sim 35$ & $5.5$ & $0.8$ & $f_{\mathrm{esc}} \propto \mathrm{M}_\mathrm{halo} (> 10^{10})$\\
		\thesanlow & $95.5$  & $2 \times 1050^3$ & $2.49 \times 10^7$ & $4.66 \times 10^6$ & $4.1$ & $\sim 35$ & $5.5$ & $0.95$ & $f_{\mathrm{esc}} \propto \mathrm{M}_\mathrm{halo} (<10^{10})$\\
		\thesansdao & $95.5$  & $2 \times 1050^3$ & $2.49 \times 10^7$ & $4.66 \times 10^6$ & $4.1$ & $\sim 35$ &  $5.5$ & 0.55 & Strong dark acoustic oscillations\\
		\\
		\thesantng & $95.5$  & $2 \times 1050^3$ & $2.49 \times 10^7$ & $4.66 \times 10^6$ & $4.1$ & $\sim 35$ &  $5.5$ & - & original TNG model\\
        \thesantngsdao & $95.5$  & $2 \times 1050^3$ & $2.49 \times 10^7$ & $4.66 \times 10^6$ & $4.1$ & $\sim 35$ &  $5.5$ & - & original TNG model + sDAO\\
		\thesannort & $95.5$  & $2 \times 1050^3$ & $2.49 \times 10^7$ & $4.66 \times 10^6$ & $4.1$ & $\sim 35$ &  $5.5$ & - & no radiation\\
		\\
		\thesandarkone & $95.5$  & $2100^3$ & $3.70 \times 10^6$ & - & $2.2$ & - &  $0.0$ & - & DM only\\
		\thesandarktwo & $95.5$  & $1050^3$ & $2.96 \times 10^7$ & - & $4.1$ & - &  $0.0$ & - & DM only\\
		\hline
	\end{tabular}
\end{table*}

\subsection{Dust creation and destruction}
\label{sec:met:dust}
We augment the IllustrisTNG galaxy formation model with a set of empirical relations that describe the production and destruction of cosmic dust. We follow the approach of \citet{McKinnon2016, McKinnon2017} and numerically treat dust as a property of the gas resolution elements. Hence, we neglect relative motion between these two components and passively advect dust across gas cells during the hydrodynamical step. Within individual resolution elements the model tracks the mass of dust in five chemical species (C, O, Mg, Si and Fe). The dust mass can increase via two processes: dust production during stellar evolution, and grain growth in the ISM. In the former, we assume part of the metals returned to the ISM condensate into dust within a single simulation timestep. This fraction depends on the stellar evolution phase --~either asymptotic giant branch (AGB) or SN~-- and on the carbon-to-oxygen ratio (C/O) of the star \citep{Dwek1998}. 

Once produced, dust is allowed to grow via deposition of gas-phase metals. We model this process computing the local instantaneous dust growth rate as
\begin{equation}
    \frac{\dd M_{\mathrm{dust}}}{\dd t} = \left( 1 - \frac{M_{\mathrm{dust}}}{M_{\mathrm{metal}}} \right) \frac{M_{\mathrm{dust}}}{\tau_\mathrm{g}}~, \qquad\qquad \text{(Growth)}
\end{equation}
where $M_{\mathrm{metal}}$ refers to gas-phase metals only, and $\tau_\mathrm{g}$ is the growth time scale, which depends on the local gas temperature and density, and is modelled to resemble conditions found in molecular clouds \citep[in particular, we are employing the same values used in][]{McKinnon2017}.

Many physical processes can destroy cosmic dust. Among others, shocks \citep[\eg][]{Seab&Shull83, Seab87, Jones+94}, sputtering \citep[both thermal and non-thermal, \eg][]{Draine&Salpeter79b, Tielens+94}, and grain-grain collisions \citep[\eg][]{Draine&Salpeter79a, Jones+96}.
We neglect the latter \citep[since it is expected to be sub-dominant, see \eg][]{Barlow78, Draine&Salpeter79a, Draine&Salpeter79b, Jones+94}, and implement the effect of the other processes by computing local instantaneous dust destruction rates as
\begin{equation}
    \frac{\dd M_{\mathrm{dust}}}{\dd t} = -\frac{M_{\mathrm{dust}}}{\tau_\mathrm{sh}} -\frac{M_{\mathrm{dust}}}{\tau_\mathrm{sp}/3}~. \qquad\qquad \text{(Destruction)}
\end{equation}
Here, $\tau_\mathrm{sh}$ and $\tau_\mathrm{sp}$ are the shock- and sputtering- driven destruction time scales, respectively. The former depends upon the local dust mass, the grain-destruction efficiency of SN shocks, the local rate of type II SN, the type II SN energy, and the typical shock velocity \citep[see][]{McKinnon2016}. In computing $\tau_\mathrm{sp}$ we assume that thermal sputtering processes dominate over non-thermal ones, hence it solely depends upon the local gas temperature and density, and the grain size (which we assume to be $a = 0.1 \, \mu \mathrm{m}$; \citealt{McKinnon2017}). Finally, dust is captured within stellar particles during star formation (a mechanism often referred to as astration). During this process, the relative importance of gas- and dust-phase metals is maintained constant. We note that the mass of metals is reduced because a fraction of it is locked up in dust grains. However, this does not have any appreciable effect on the cooling/star formation rates, compared to the original IllustrisTNG model because the amount of dust is low, at least in the redshift range we consider here.

\subsection{Initial conditions}
The initial conditions employ a method in which the initial Fourier mode amplitudes are fixed to the ensemble average power spectrum \citep{Angulo&Pontzen2016} to suppress variance. During the production of the initial conditions, each power spectrum mode $P (k)$ is usually sampled from a Gaussian distribution centred on the expectation value for the given cosmological model with true Gaussian fluctuations $\hat{P} (k)$. In the \textit{fixed} approach, we set $P (k) = \hat{P} (k)$ for each (discrete) $k$ value sampled, eliminating the variance on linear scales. This renders the initial conditions non-Gaussian, but the latter has negligible effects on the results of the simulations \citep[see \eg][]{Angulo&Pontzen2016, Villaescusa-Navarro+2018, Anderson+2019, Klypin+2020}. Using this approach ensures we obtain the optimal (\ie most-representative of the adopted cosmology) initial conditions. A notable consequence is that the halo mass function is ensured to match the ensemble average from many non-fixed realisations.

On a practical level, the initial conditions are produced with the \textsc{Gadget-4} code \citep{Gadget4} using second order Lagrangian perturbation theory, at the initial redshift of $z_\mathrm{in} = 49$. We employ a \citet{Planck2015_cosmo} cosmology (more precisely, the one obtained from their \texttt{TT,TE,EE+lowP+lensing+BAO+JLA+H$_0$} dataset), \ie $H_0 = 100h$ with $h=0.6774$, $\Omega_\mathrm{m} = 0.3089$, $\Omega_\Lambda = 0.6911$, $\Omega_\mathrm{b} = 0.0486$, $\sigma_8 = 0.8159$, and $n_s = 0.9667$, where all the symbols have the usual meaning. The gas perfectly follows the DM distribution in the initial conditions and is assumed to have primordial composition with hydrogen and helium mass fractions of $X=0.76$ and $Y=0.24$, respectively. 

\subsection{Simulations}
The full \thesan simulation set and the associated parameters  are catalogued in Table~\ref{table:simulations}. All our simulation runs follow the evolution of a cubic patch of the universe with linear comoving size $\Lbox = 95.5 \, \mathrm{cMpc}$. We seek to simulate all galaxies that are thought to be important for the reionization process. Therefore, we aim to model the so-called atomic cooling halos, where (i) the virial temperature is sufficiently large ($T_\mathrm{vir} \gtrsim 10^4$ K) to allow primordial cooling, and (ii) the gravitational potential is sufficiently large to retain photoheated gas. These halos have a typical mass at the EoR ($z=6-10$) of $\Mh \sim 10^8 \, \hMsol$ \citep[see \eg][]{Bromm&Yoshida2011}. We consider a DM halo resolved if it contains $\sim 50$ particles\footnote{We note that this is the minimum resolution necessary to get roughly converged star formation rates and gas mass fractions in these low mass haloes.}, which sets our fiducial (highest) mass resolution to $m_\mathrm{DM} = 3.12 \times 10^6 \, \mathrm{M}_\odot$, and $m_\mathrm{gas} = 5.82 \times 10^5 \, \mathrm{M}_\odot$ for DM and gas, respectively. This sets the total number of dark matter and (initial) gas particles to $N_\mathrm{particles} = 2100^3$ each. The gravitational softening length for the star and dark matter particles is set to $2.2$ \ckpc. The gas cells are adaptively softened according to the cell radius with a minimum value set to $2.2$ \ckpc. The gas cells are (de-)refined so that the gas mass in each cell is within a factor of two of the target gas mass ($m_\mathrm{gas}$). This implies that the gas cells in the highest density regions will have smaller sizes than the minimum softening length. In the highest resolution \thesanone simulation the minimum cell radius at $z=5.5$ is $\sim 10$ pc, which is the smallest scale over which the gas hydrodynamics and baryonic processes are resolved. 

The motivation for the choice of these softening lengths comes from considerations outlined in \citet{Ludlow2019a, Ludlow2019b} and \citet{Ludlow2020} which are improvements to the previous studies on this topic by \citet{Power2003}. Their detailed convergence studies showed that the softening lengths should be  sufficiently large to suppress 2-body scattering as much as possible, but sufficiently small so that gravitational forces are unbiased on the relevant spatial scales. The suggested optimal softening is $\varepsilon_\mathrm{opt} \sim 0.05 \, \Lbox / N_\mathrm{particles}$ which results in a value of about $2.2$ \ckpc \, for \thesanone. We also note that the physical softening length is always larger than the minimum value required to resolve an escape velocity of $10 \, \mathrm{km \, s}^{-1}$, at all relevant redshifts. This is necessary to ensure that the effects of photoionization heating associated with reionization are not artificially suppressed. 

We also include one additional parameter, $f_\mathrm{esc}$, that mimics the absorption of LyC photons happening below the grid scale of the simulation. This parameter is tuned such that the simulated reionization histories approximately match the observed neutral fraction evolution in the Universe \citep[see for example, ][]{Greig2017}. 

\begin{figure}
	\includegraphics[width=0.99\columnwidth]{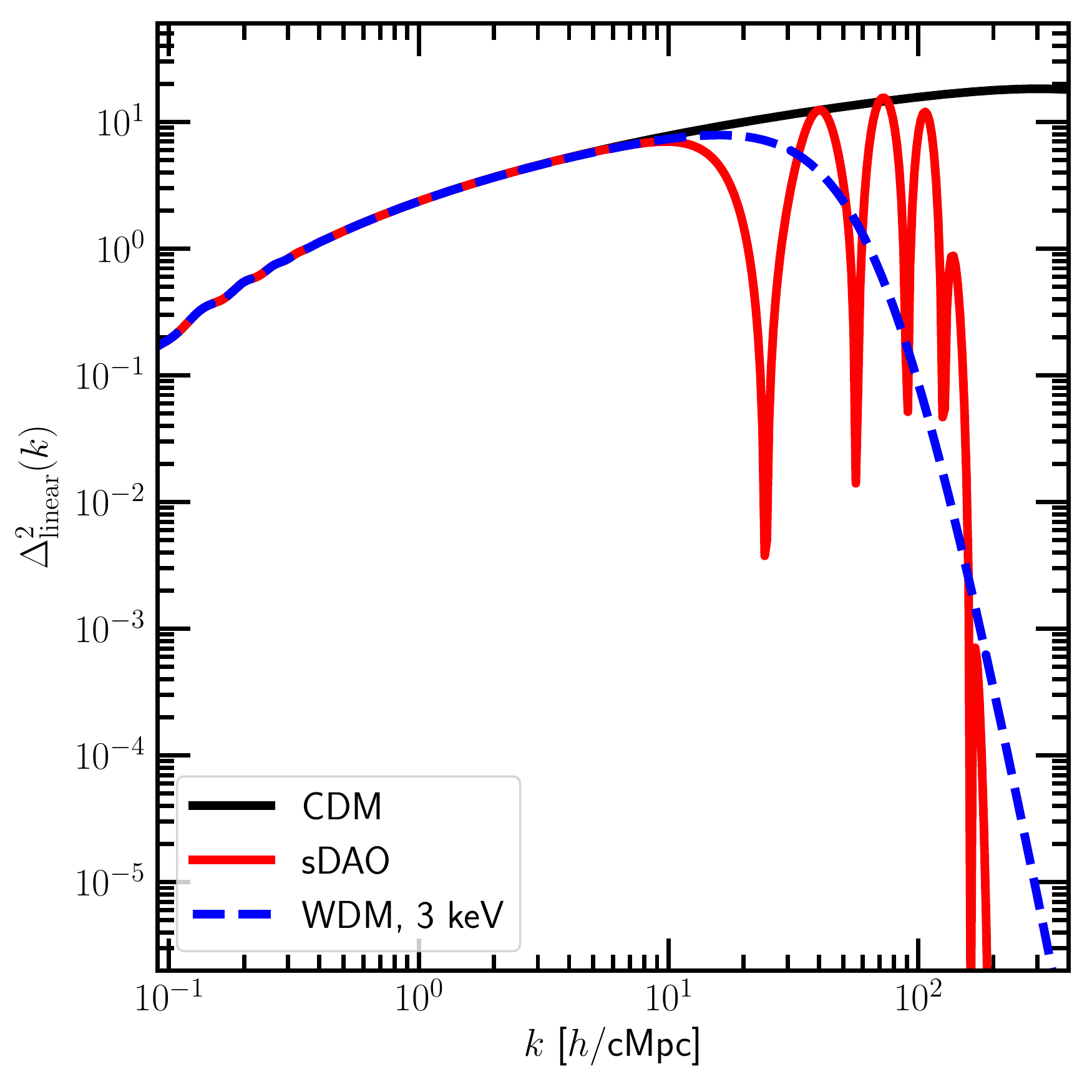}
    \caption{The linear initial power spectra for the CDM (black curve) and strong dark acoustic oscillation (sDAO; red curve) models as a function of comoving wavenumber as used in our simulations. For reference we also plot the power spectrum for a WDM model ($m_\mathrm{DM} = 3 \, \mathrm{keV}$).}
    \label{fig:ics}
\end{figure}

The  \thesan simulation suite also features a set of medium-resolution runs designed to investigate the changes to reionization induced by different escape fractions from galaxies, varying dark matter models, the impact of assuming a constant radiation background in galaxy formation simulations and numerical convergence. These additional runs employ identical initial conditions, but the mass resolution is lower by a factor of $8$ and the softening length is increased by about a factor of two. \thesantwo is a medium resolution run that uses the same fiducial model and escape fraction as \thesanone. \thesanwc is the same as \thesantwo except for the slightly higher escape fraction, which tries to compensate for the lower star formation rate in the medium resolution runs (see Section~\ref{sec:results} for more details) such that the total integrated number of photons emitted in \thesanone and \thesanwc runs are the same. The \thesanhigh and \thesanlow simulations have been specifically designed to understand whether the high \citep{Naidu2020} or low \citep{Finkelstein2019} mass galaxies dominate the reionization photon budget.  These simulations use a halo mass dependent escape fraction with only halos above (\thesanhigh)/below (\thesanlow) $10^{10}\, \Msun$ contributing to the reionization process. On a practical level, this is achieved by tagging on to the in-built friends-of-friends (FoF; \citealt{Springel2005}) algorithm to calculate the halo mass of the group in which the stellar particles reside and deciding the escape fraction of photons accordingly. 

\thesansdao aims to probe the impact of assuming non-standard DM models on the reionization process. We particularly focus on models that cutoff the linear matter power spectrum at small scales due to collisional damping caused by interactions between DM and relativistic particles in the early Universe resulting in Dark Acoustic Oscillations (DAOs) \citep{DAO}. We note that the only difference between \thesantwo and this simulation lies in the linear matter power spectrum which is quantified by the transfer function, defined as the square root of the ratio between the power spectrum in the DAO model and the canonical CDM model. We use the parametrisation outlined in Equation~$3$ of \citet{Bohr2020}, with $k_\mathrm{peak} = 40 \hMpc$ and $h_\mathrm{peak}=1$ to compute the transfer function. $k_\mathrm{peak}$ determines the position of the first DAO peak and describes the position of the cutoff, while $h_\mathrm{peak}$ sets the amplitude of the DAO oscillations. Figure~\ref{fig:ics} shows the linear matter power spectrum of the CDM (black curve) and  strong DAO (sDAO; red curve) models used in our simulation suite (the plot also shows the same for a Warm Dark Matter model, blue dashed curve, with $m_\mathrm{DM} = 3$ keV for comparison).

\thesantng and \thesantngsdao use the same ICs as \thesantwo and \thesansdao respectively, but employ the original IllustrisTNG model including the assumption of an instantaneous reionization process and a spatially uniform ultra-violet background \citep{CAFG2009} below the reionization redshift. \thesannort does not consider any radiation background and is designed to understand the role of phtoionization feedback due to reionization. Finally, we also run two DM only simulations at fiducial (\thesandarkone) and medium (\thesandarktwo) resolutions. They are completed down to $z=0$, allowing us to trace the local progenitors of the simulated high-redshift galaxies. This exhaustive set of simulations allows us to probe the reionization process and the impact of various physical assumptions in greater detail.

\begin{figure*}
	\includegraphics[width=1.97\columnwidth]{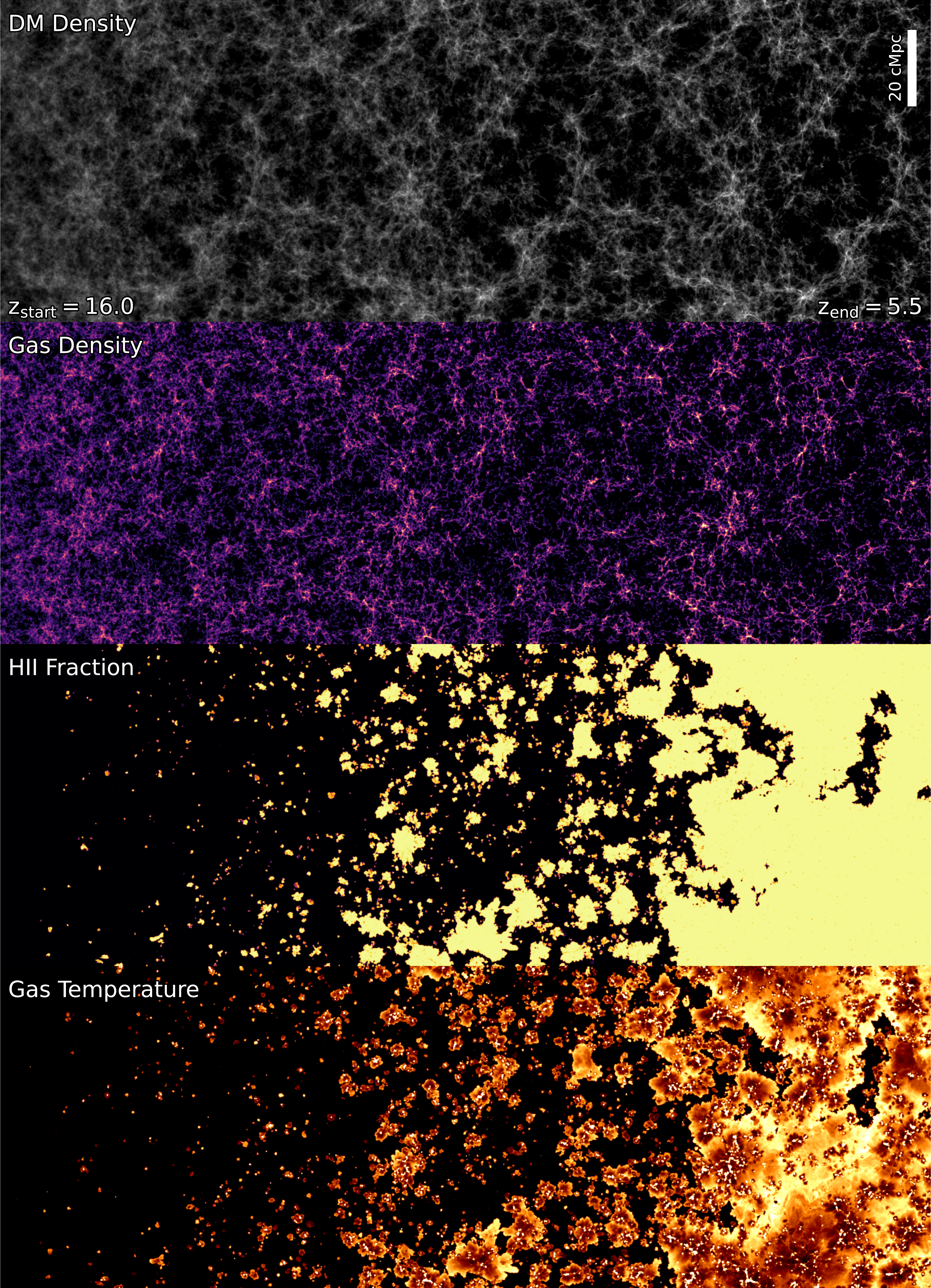}
    \caption{Mock light cones showing the evolution of DM density, gas density, ionized fraction, and temperature of the intergalactic medium. The \hii regions start out small at high redshifts and then get bigger and merge with each other as reionization progresses, eventually ionizing all the IGM in the Universe.}
    \label{fig:evolve}
\end{figure*}

\begin{figure*}
	\includegraphics[width=0.99\textwidth]{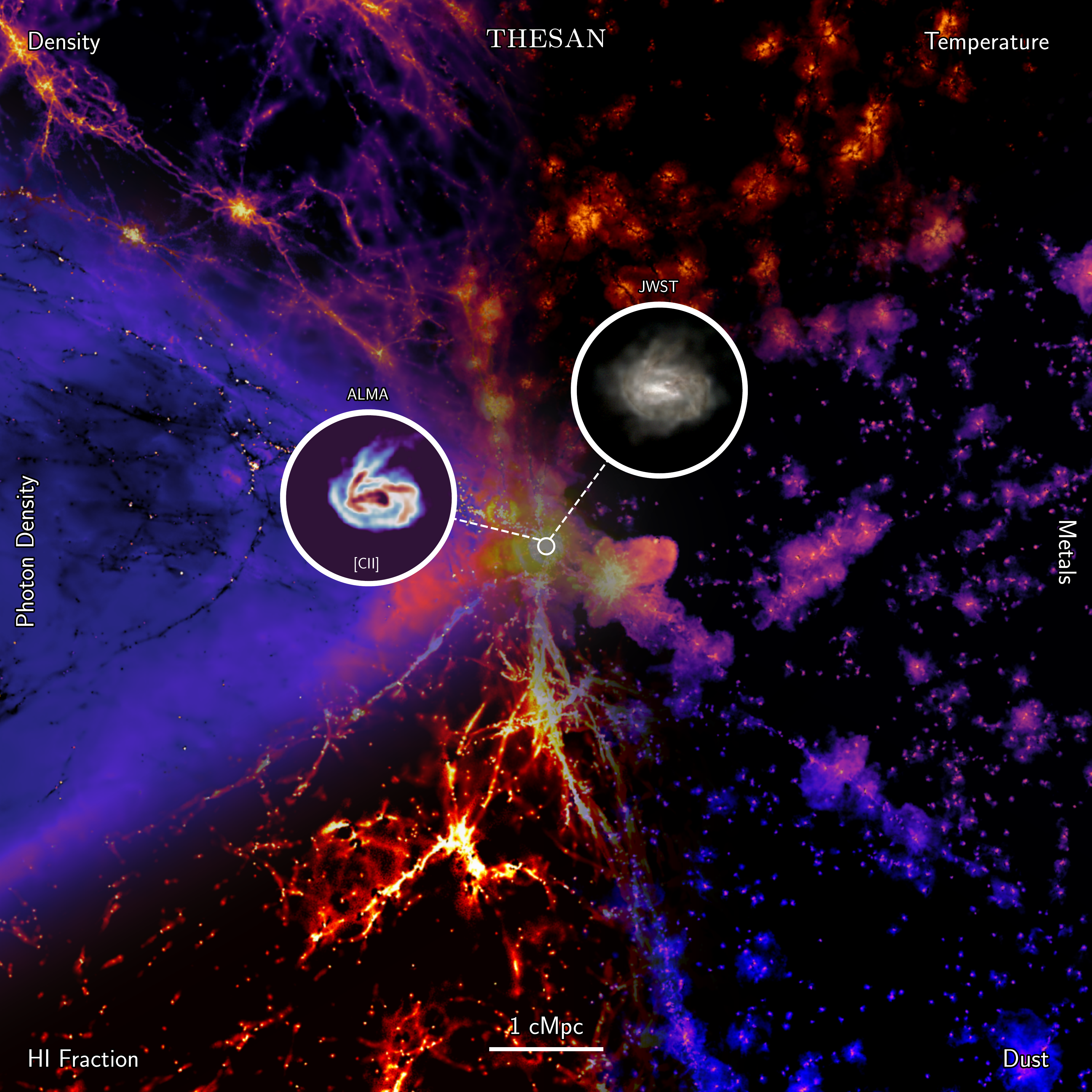}
    \caption{A representation of a sample of quantities predicted by the \thesan simulations. From top left in clockwise order the image shows the gas density, temperature, metal and dust distribution, neutral hydrogen fraction, and photon density. The central zoom-in regions show mock \jwst (F277W, F356W, F444W) and \alma ([CII]) images of the largest galaxy present in the simulation at $z=5.5$. The large boxsize and high spatial resolution coupled with a well tested galaxy formation model allows us to precisely capture both the large scale nature of the patchy reionization process and at the same time self-consistently predict the properties of the sources responsible for it.}
    \label{fig:one}
\end{figure*}

\subsection{Code Deployment}
The Flagship simulation was performed on the SuperMUC-NG supercomputer at Leibniz-Rechenzentrum. The code was deployed on $57\,600$ cores, the minimum amount necessary to hold the simulation memory. The simulation was run between August 2020 and March 2021 and used a total of about $28$ million core hours. We note that in order to reduce the memory requirements the simulation used a mixed precision scheme which employed double precision for a small fraction of the variables and switched all other variables to single precision. It was necessary to retain the particle coordinates in double precision to ensure that particle positions did not overlap. Additionally, the photon number densities and fluxes also used double precision in order to get stable solutions during the thermochemistry step. We have extensively tested this scheme and are
confident that this does not introduce any spurious effects in the simulations. Additionally, a new feature of MPI-3, which allows the allocation of shared memory that can be jointly accessed by the MPI ranks residing on the same shared memory node is used for storing data that is equal on all MPI-ranks like top-level tree and large look-up tables. This allows for storing data only once per compute node, and not for all MPI ranks separately.

\subsection{Data Products}

A total of 80 snapshots were written, every 10.9 Myr, from $z = 20$ down to $z = 5.5$, each one of size $\sim 1.9$ TB. The dark matter halos are identified using the friends-of-friends  (FOF) algorithm \citep{Davis1985} using a linking length of 0.2 times the initial mean inter-particle distance. Stellar particles and gas cells are attached to these FOF primaries in a secondary linking stage. The SUBFIND algorithm \citep[first described in ][]{Springel2001} is then used to identify gravitationally bound structures. These FOF and SUBFIND catalogues accompany each snapshot output and contain a wide array of information about the detected DM halos and the gas  and stellar properties of the halos.  Additionally, we also output high time cadence Cartesian data which grids the simulation data  onto a regular Cartesian grid using a first order Particle-In-Cell approach. \thesanone uses a $1024^3$ grid while the medium resolution runs use a $512^3$ grid, setting the cell sizes to $~\sim 93$ \ckpc\xspace and $\sim 186$ \ckpc\xspace,  respectively. While most of the quantities like the gas and stellar density, ionization fraction and temperature are binned on a regular spatial grid, observables like the 21\,cm emission (obtained from the neutral hydrogen density) and \lyalpha luminosities are binned in redshift space which accounts for Doppler shifting due to the peculiar velocities of the galaxies (i.e. individual gas cells).
This redshift-space binning is carried out assuming the observed direction is aligned with the $+z$-axis.
The Cartesian outputs are written every $\sim 2.8$\,Myrs amounting to $400$ snapshots over the duration of the simulation. These lower-resolution higher-time-cadence outputs are ideal for studying the large scale properties of the reionization process. The full list of Cartesian output quantities and the structure of the files will be outlined during the public release of the \thesan data.

\begin{figure}
	\includegraphics[width=0.99\columnwidth]{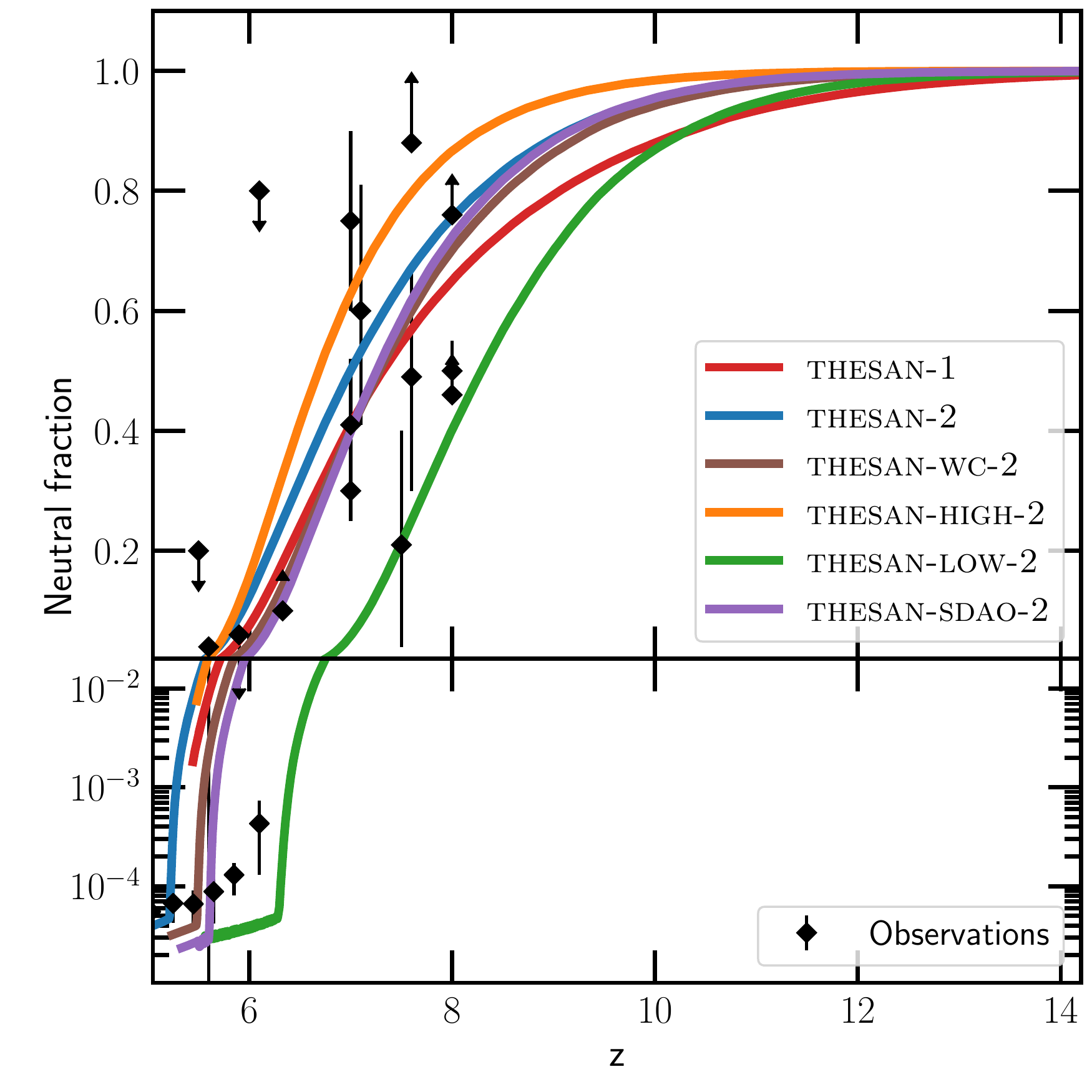}
    \caption{Evolution of the volume weighted neutral fraction as a function of redshift in our various simulations as indicated. The observational estimates from \citet{Fan2006}, \citet{McGreer2011, McGreer2015}, \citet{Ono2012}, \citet{Schroeder2013}, \citet{Choudhury2015}, \citet{Greig2017}, \citet{Mason2018, Mason2019},  \citet{Greig2019}, \citet{Hoag2019} and \citet{Jung2020} are shown as black diamonds. All simulations, except \thesanlow, follow a late reionization scenario and contain relatively large neutral regions even below $z=6$. }
    \label{fig:all}
\end{figure}

\section{Results}
\label{sec:results}

Figure~\ref{fig:evolve} shows a mock lightcone like visualization illustrating the evolution of the DM density (top row), gas density (second row), hydrogen ionization fraction (third row), and  gas temperature (fourth row) in the \thesanone simulation. The DM starts out fairly uniform at $z=16$ and builds up structures progressing to lower redshifts.
Gas then collapses in the dark matter halos by radiating away its energy. The dense gas cools and forms the very first stars and galaxies in the Universe. 

These early stars and galaxies emit large amounts of LyC radiation, which ionizes and heats up the cold neutral IGM. The initial \hii regions around the early galaxies are quite small. This is expected as the amount of stars formed in these early low mass halos is quite small. As time progresses the ionized regions begin to become larger as the galaxies become bigger and the star formation rate increases. Eventually, multiple nearby \hii regions overlap to form larger and larger ionized bubbles until
all the IGM in the Universe is ionized. The temperature structure mimics the evolution of the ionization fractions with the ionized gas photoheated to a temperature of about $\sim 10^4$ K. The additional small fluctuations (which are absent in the ionization field) in the temperature structure are caused by the difference in time since reionization (and hence the amount of cooling that can happen afterwards) and by the different speed of the ionization fronts (I-fronts) at different gas densities with slower speeds at higher gas densities and vice-versa \citep{Daloisio2019}.

\begin{figure}
	\includegraphics[width=0.99\columnwidth]{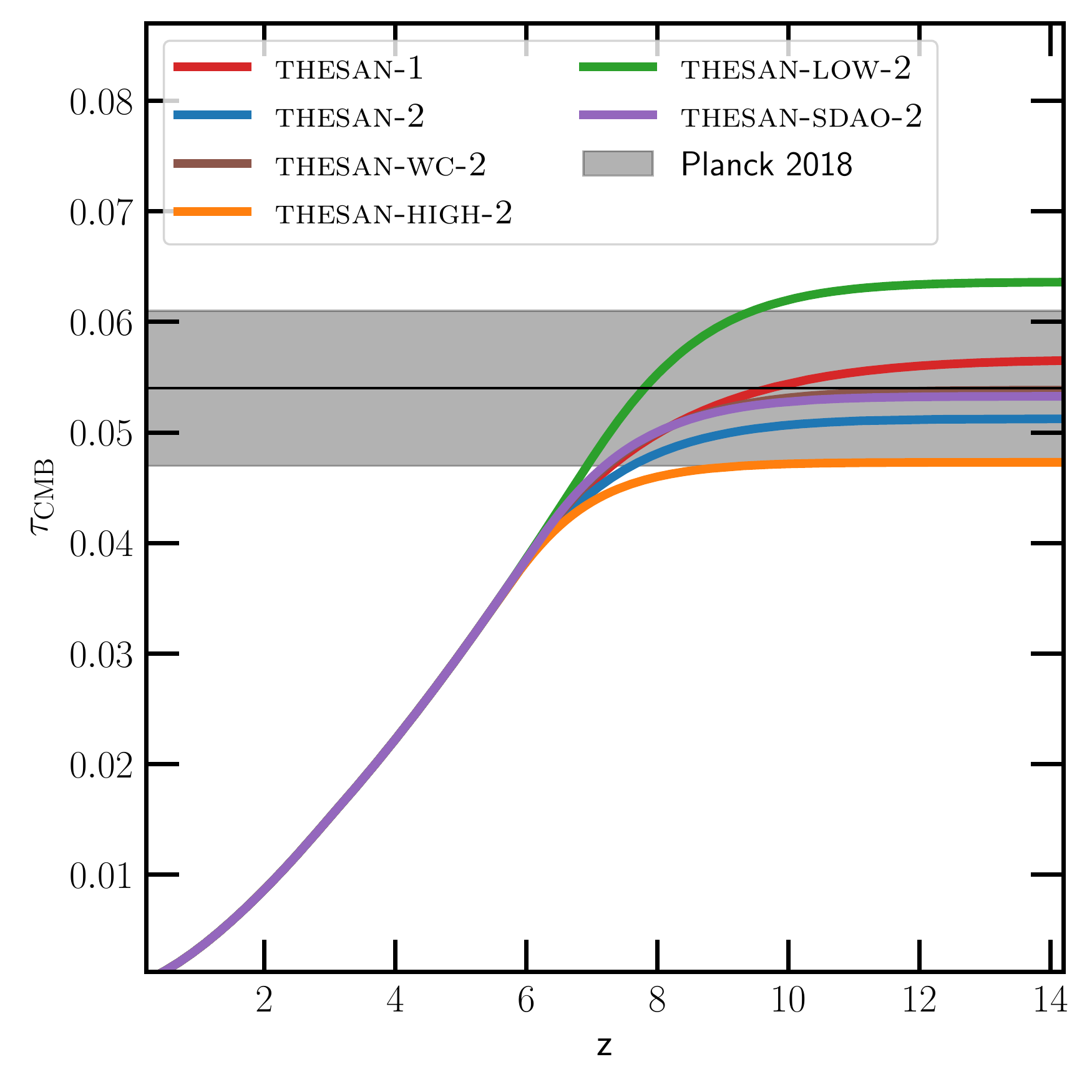}
    \caption{Optical depth to the CMB compared to the \citet{Planck2018} estimates. The simulations which only reionize completely below $z=6$ show optical depths that are consistent with the observational estimates.  \thesanlow, on the other hand, with its early reionization history produces a slightly larger value.}
    \label{fig:cmb}
\end{figure}

\begin{figure}
	\includegraphics[width=0.99\columnwidth]{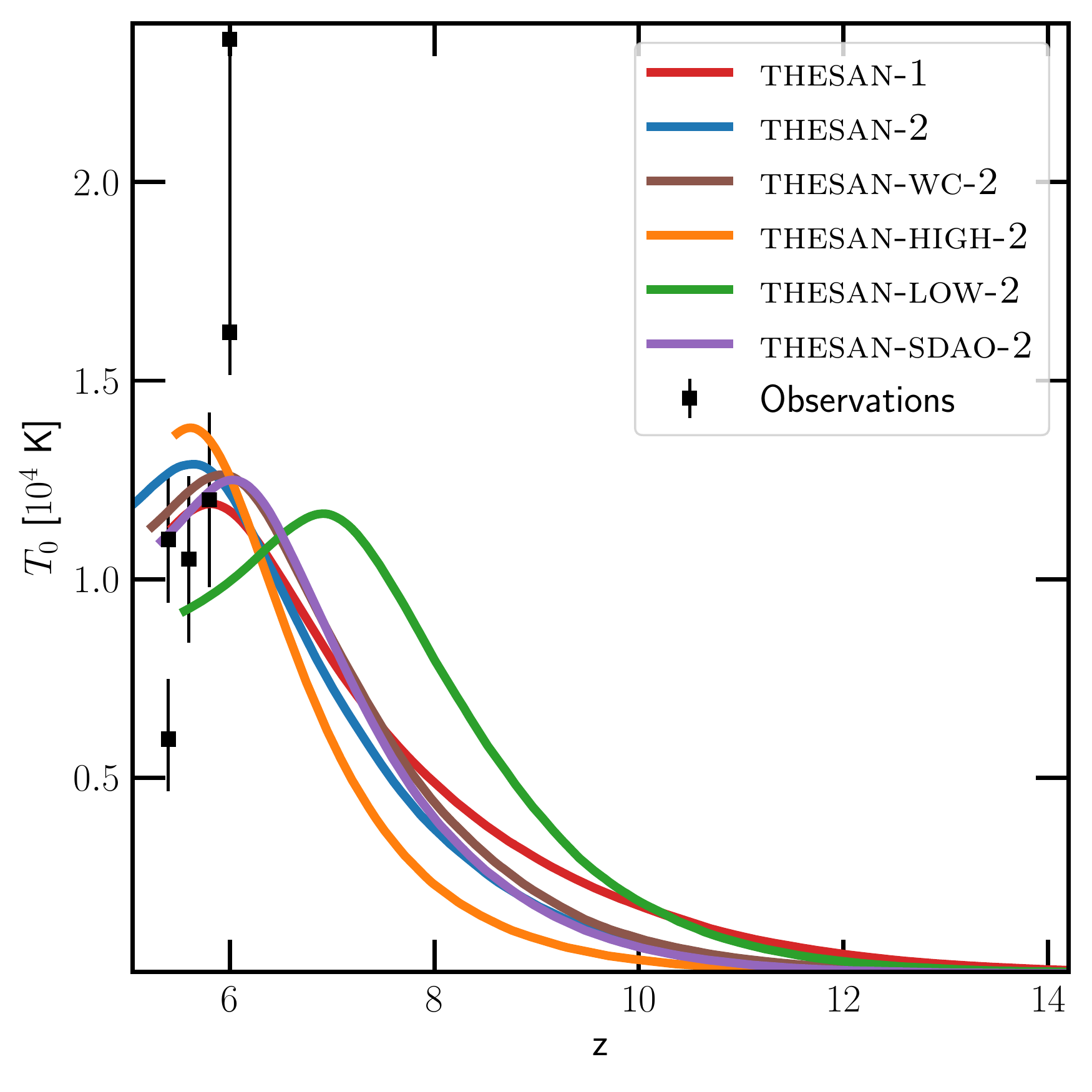}
    \caption{Evolution of the IGM temperature ($T_0$) at mean density for the various simulations compared to observational estimates from \citet{Bolton2010, Bolton2012}, \citet{Walther2019} and \citet{Gaikwad2020}. Simulations with shorter reionization histories have faster I-fronts, which produce hotter post I-front tempertures leading to a slightly higher maximum $T_0$. }
    \label{fig:T0}
\end{figure}

Figure~\ref{fig:one} zooms in on a $10 \, \cMpc$ region around the most massive galaxy in the simulation volume at $z=5.5$ and presents a visual overview of a sample of properties predicted self-consistently by the \thesan simulations. It shows the gas density (top left), temperature (top right), metal mass (right), dust mass (bottom right), the neutral hydrogen fraction (bottom left) and the ionizing photon density in the $13.6 - 24.6$ eV band (left). The temperature peaks inside massive DM halos as the infalling gas gets shock heated to the virial temperature of the halo \citep{Dekel2006}. While most of the dust and metal mass is concentrated inside galaxies,  a non-negligible fraction resides in the CGM and even  the IGM. These are most likely transported out to these distances by stellar and black-hole driven galactic outflows. Although most of the hydrogen in the Universe is ionized by $z=5.5$, the gas in the high density filaments and nodes is self-shielded against the background radiation field and remain largely neutral. These neutral regions are observed as Lyman limit and damped \lyalpha systems in the \lyalpha forest \citep{Prochaska1999}. This is reflected in the photon density with the low density ionized gas exhibiting a significant radiation field, while the high density filaments remain largely photon free. The insets show mock \jwst and \alma images of the central galaxy. These images were generated with \textsc{skirt} \citep[last described in ][]{skirt}, using the methodology outlined in \citet{Vogelsberger2020}. The mock \jwst image is based on F277W, F356W and F444W NIRCam wide filters. It covers a $\sim 12.5 \, \kpc$ (diameter) field of view. The \alma image shows predictions for the expected emission from the singly ionized carbon line ([CII]) at $158 \mu\mathrm{m}$. The simulated emission is convolved with a point spread function of about $\sim 17$ mas, which is the expected resolution of the telescope in its highest resolution configuration\footnote{\url{https://almascience.nrao.edu/about-alma/alma-basics}}. 

These powerful visualizations demonstrate how
the relatively large box size and high
spatial resolution coupled with a well tested galaxy formation model allows us to precisely capture both the large scale nature of the patchy reionization process and the same time self-consistently predict the properties of the sources (galaxies and black holes) responsible for it.

\subsection{Reionization history}
\label{sec:reion}
For a more quantitative analysis of the simulations we begin with Figure~\ref{fig:all}, which shows the evolution of the volume weighted neutral fraction as a function of redshift ($z$), for the \thesanone (red curve), \thesantwo (blue curve), \thesanwc (brown curve), \thesanhigh (orange curve), \thesanlow (green curve) and \thesansdao (purple curve) runs. For comparison we show a variety of recent observational estimates (black diamonds) derived from the effective optical depth of the \lyalpha forest \citep{Fan2006}, \lyalpha dark pixel statistics  \citep{McGreer2011, McGreer2015}, statistics of \lyalpha emitting systems \citep{Ono2012, Choudhury2015, Mason2018, Mason2019, Hoag2019, Jung2020} and \lyalpha damping wings in the near-zones of high-redshift quasars \citep{Schroeder2013, Greig2017, Greig2019} . All simulations roughly match the neutral fraction evolution. \thesanone shows a very extended reionization history. This is because a lot of star formation occurs in halos with $\Mh \lesssim 10^9 \Msun$ at high redshifts. These low mass halos ionize their immediate surroundings producing a non-negligible ionization fraction ($\gtrsim 0.1$) even at very high redshifts ($z>10$). These low mass halos are, however,  unresolved in the medium resolution runs and therefore the early ionization of the IGM is hindered. This forcibly reduces the duration of reionization in all the medium resolution runs. A more thorough discussion of these effects will be taken up Section~\ref{sec:galaxies}.

We also note that the \thesanone, \thesantwo and \thesanhigh runs are unable to fully reionize the Universe by $z=5.5$ with the neutral fraction reaching values of about $10^{-3}$ for \thesanone and $10^{-2}$ for the other two. This corresponds to about $0.1\%$ and $1\%$ of the total simulation volume being neutral at the end of the simulation, respectively. These neutral regions below $z=6$ can be seen clearly in Figure~\ref{fig:evolve} in both the ionization and temperature fields. We extend the \thesantwo simulation down to $z=5$ and the neutral region vanishes by $z=5.1$. We believe that the other two simulations, had we extended them down to $z=5$, would have been fully reionized as well. These results agree with `late' reionization models invoked to explain the observed long troughs in the \lyalpha forest in the spectra of high-redshift quasars \citep{Kulkarni2019, Keating2020}. The impact of these neutral islands on the \lyalpha forest will be investigated in detail in an accompanying paper \citep{GaraldiThesan}.

\begin{figure}
	\includegraphics[width=0.99\columnwidth]{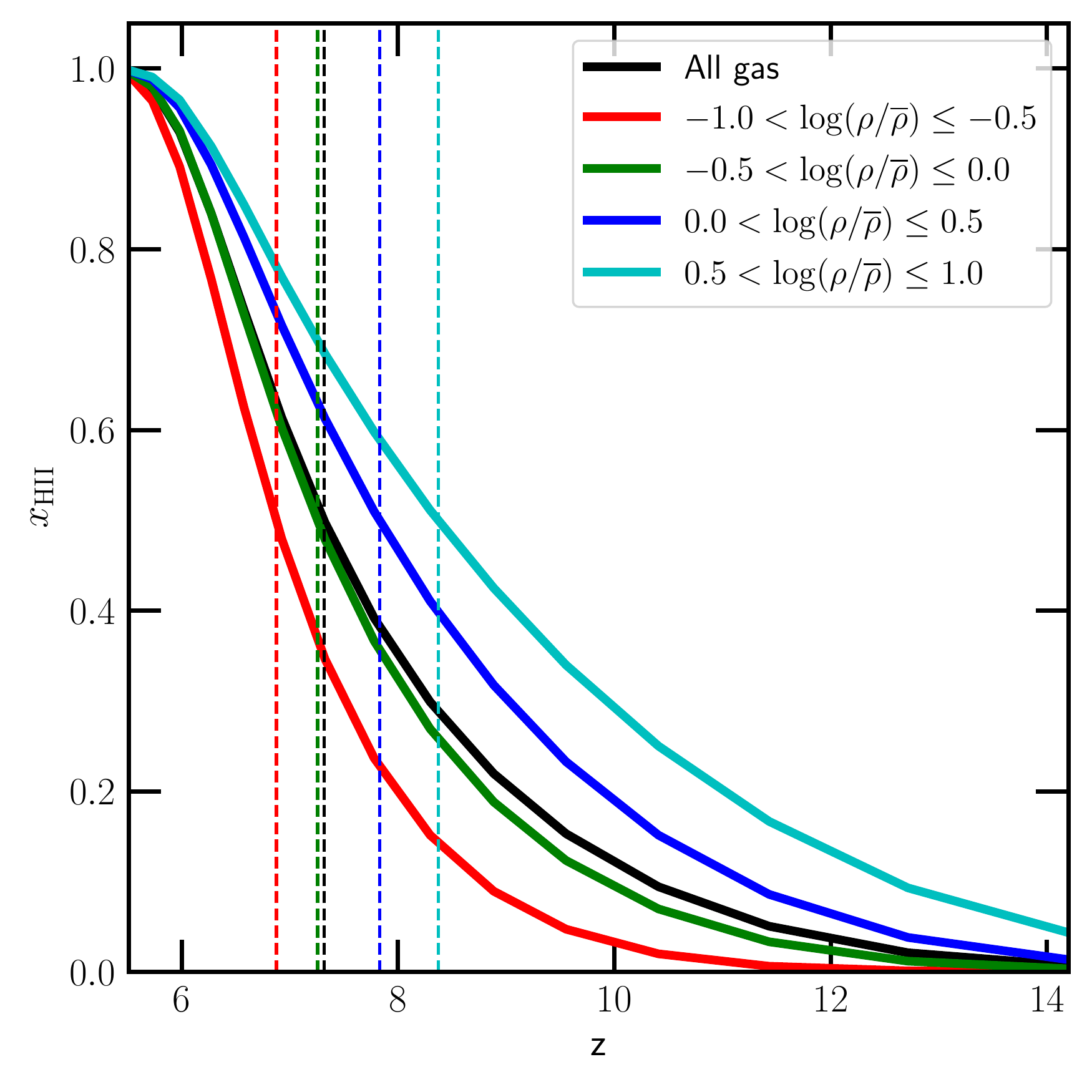}
    \caption{Ionization histories at different gas densities in the \thesanone simulation. The mean redshift of reionization (defined as the redshift at which $50\%$ of the IGM is ionized; vertical dashed lines) can vary by about $\Delta z\simeq 1$, with overdense gas reionizing earlier than the low density voids.}
    \label{fig:diffion}
\end{figure}

\begin{figure*}
	\includegraphics[width=0.99\textwidth]{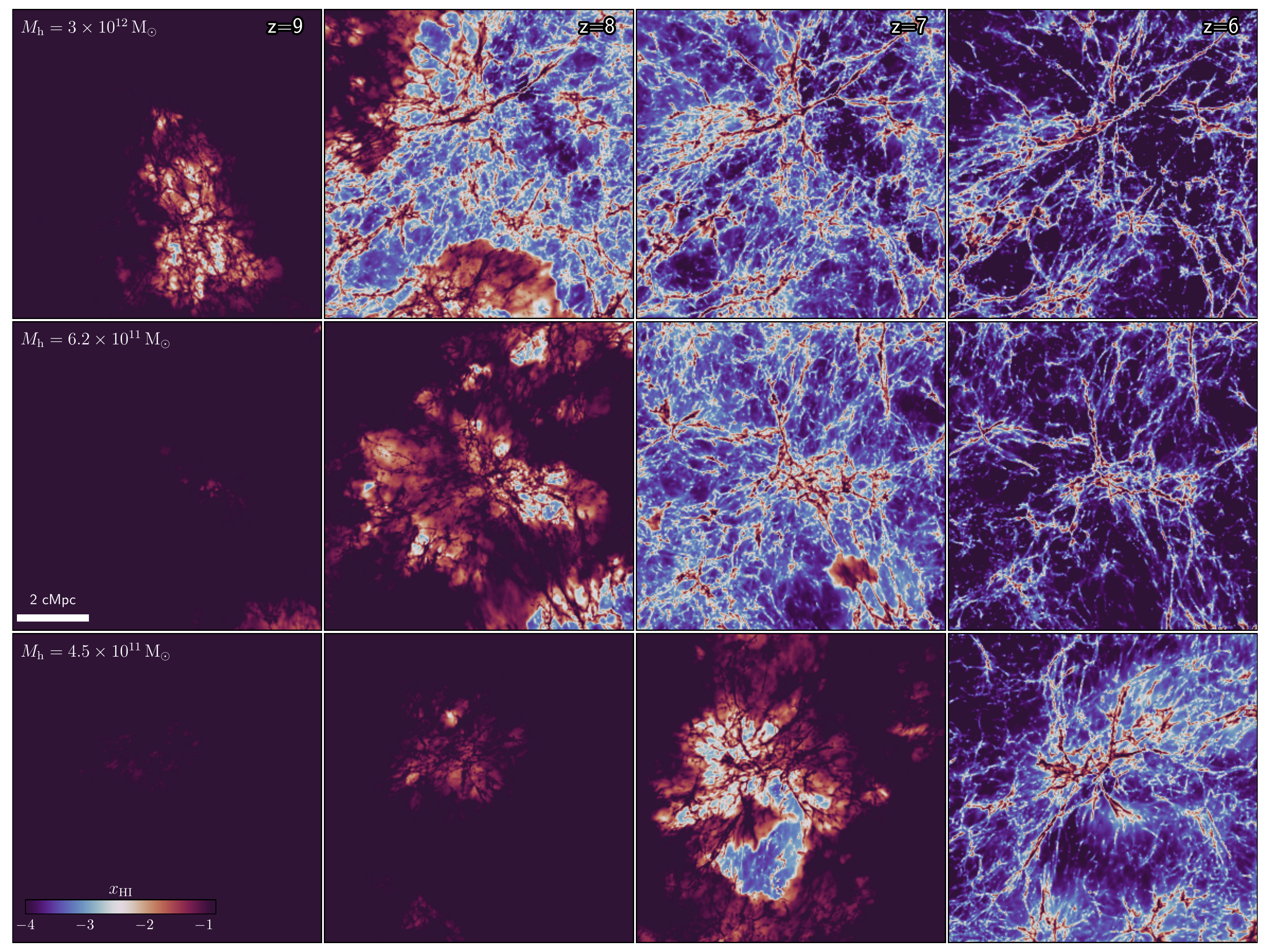}
    \caption{A visualisation showing the evolution of the neutral hydrogen fraction around three relatively large halos of mass \mbox{$M_\mathrm{h}=3\times 10^{12}\,\mathrm{M_\odot}$} (top panel), \mbox{$M_\mathrm{h}=6.2\times 10^{11}\,\mathrm{M_\odot}$} (middle panel) and \mbox{$M_\mathrm{h}=4.5\times 10^{11}\,\mathrm{M_\odot}$} (bottom panel). We note that this is the mass of the halo at $z=5.5$. Ionization fronts originate in galaxies and travel relatively slowly in the overdense regions around them, before sweeping through at very high speeds in the low density voids.   }
    \label{fig:track}
\end{figure*}

\begin{figure*}
	\includegraphics[width=0.99\textwidth]{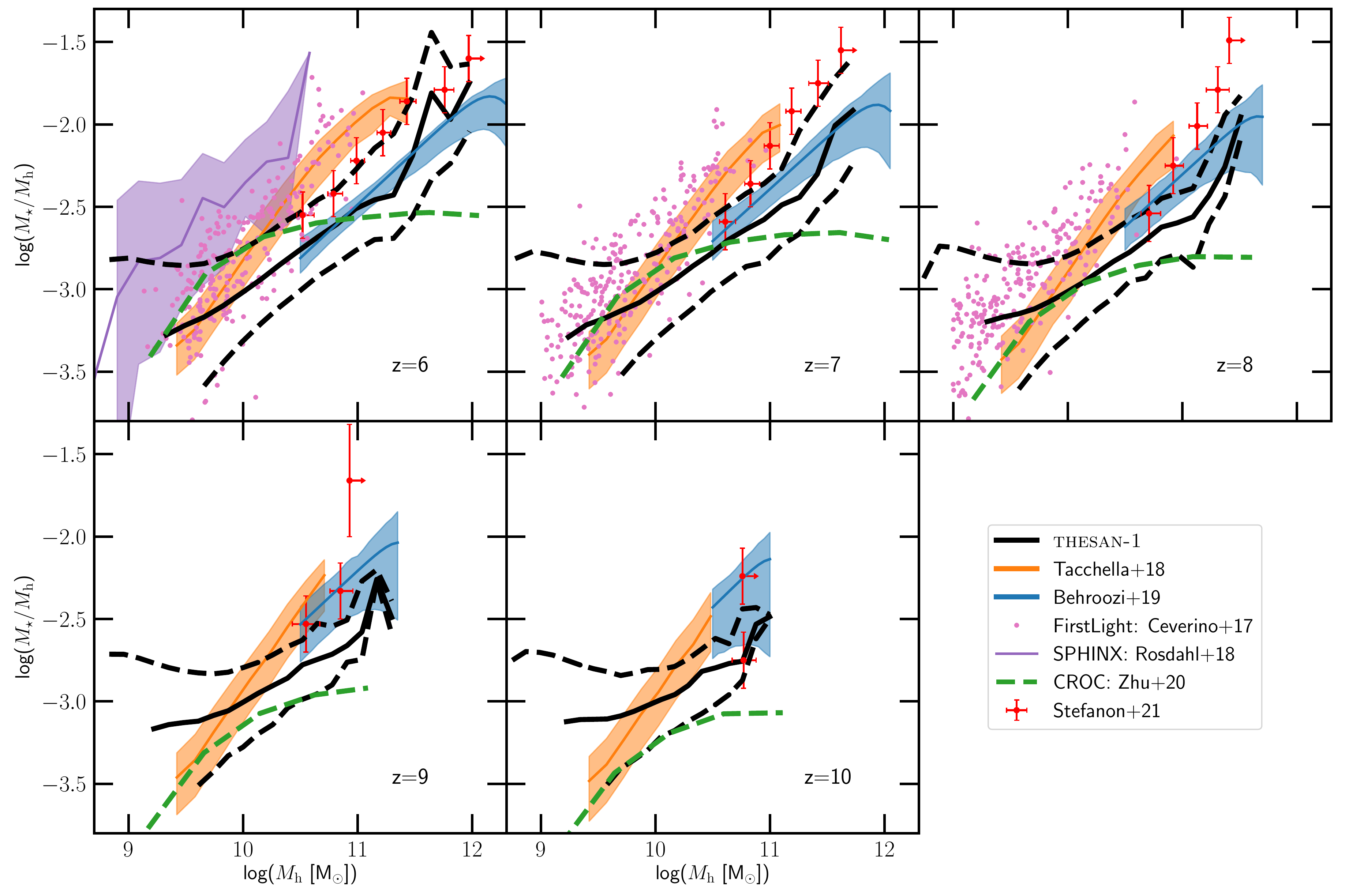}
    \caption{The stellar to halo mass relation (black curve) and its the $10$ and $90$ percentile distribution (dashed curves) of the galaxies in the \thesanone simulation (black curves) at $z=6$ (top left panel), $z=7$ (top middle), $z=8$ (top right), $z=9$ (bottom left) and $z=10$ (bottom middle). For comparison we show results from the abundance matching estimates from \citet{Behroozi2019} and \citet{Stefanon2021}, and a semi-empircal estimate from \citet{Tacchella2018}. We also show the data from other major reionization simulation efforts such as \textsc{sphinx} \citep{Rosdahl2018}, FirstLight \citep{Ceverino2017} and \textsc{croc} \citep{Gnedin2014, Zhu2020}. The \thesan results most closely match the estimated SHMR from \citet{Behroozi2019}  and are sightly below the estimates from \citet{Stefanon2021}. The \textsc{sphinx} simulations show a higher baryon conversion efficiency, while the \textsc{croc} simulations have a lower baryon conversion efficiency in high mass halos.}
    \label{fig:smhm}
\end{figure*}

Figure~\ref{fig:cmb} shows the electron-scattering optical depth of the cosmic microwave background defined as
\begin{equation}
\tau_\mathrm{CMB} (z) = c  \sigma_\mathrm{T} \int_0^{z} \frac{n_e (z)}{(1+z) H(z)} \mathrm{d}z \, ,
\end{equation} 
where $n_e$ is the electron density and $\sigma_\mathrm{T}$ is the Thompson scattering cross-section. The electron densities are self-consistently modelled by the simulations up to $z=5.5$. From $z=5.5 \, \mathrm{to} \, 3$, $n_e/n_\mathrm{H} = 1.08$, assuming full hydrogen ionization and singly ionized helium. Below $z=3$, $n_e/n_\mathrm{H} = 1.158$, assuming full ionization of both hydrogen and helium. Almost all the simulations fall within the observational range implying that reionization happens more or less at the expected redshift. \thesanlow is the only simulation that produces a slightly larger optical depth, because reionization is fully completed by $z\sim6.3$. All the `late' reionization models, on the other hand, are compatible with the \citet{Planck2018} results. Figure~\ref{fig:T0} shows the temperature of the IGM at mean density ($T_0$) as a function of redshift. The gas starts out as cold and neutral, and as the ionized fraction increases, more and more of the gas gets photoheated to about $10^4$ K,  eventually encompassing the whole Universe. \thesanone, which shows the most extended reionization history, also shows the smallest maximum \tzero, while the simulations with a shorter duration of reionization generally show larger maximum \tzero's, although this difference is quite small (only about $2000$ K). This difference arises because the simulations with a shorter reionization history, have necessarily faster ionization fronts (I-fronts) which in turn leads to hotter post I-front temperatures \citep{Daloisio2019}.\footnote{ We note that the speeds of ionization fronts can be different even if the duration of reionization is the same if they take place at different redshifts due to the non-linear dependence on redshift and time.}

Finally, Figure~\ref{fig:diffion} shows the evolution of the ionization fraction as a function of redshift for gas at different densities in the \thesanone simulation. At a particular redshift, the over-dense regions ($\rho/\bar{\rho} \gtrsim 1$) are always more ionized than the low-density regions such as voids. This is because the I-fronts originate from galaxies that reside in over-dense regions and they gradually progress through the surrounding high-density gas and into the low-density regions. This leads to a large difference in $z_\mathrm{r}$ (defined as the redshift at which the neutral fraction is 0.5) with the highly overdense gas ($0.5 < \mathrm{log}(\rho/\bar{\rho}) \leq 1.0$) having the midpoint of reionization as early as $z_\mathrm{r} \sim 8.2$ and the very low density ($-1.0 < \mathrm{log}(\rho/\bar{\rho}) \leq -0.5$) gas only reaching the midpoint at $z_\mathrm{r} \sim 7$ (see also figure 1 in \citet{GaraldiThesan}. and the discussion therein). Although the gas around galaxies experience a relatively large photonionization rate, the higher density leads to smaller Str\"omgren radii and a slower I-front expansion because its speed is inversely proportional to the  density of the gas. This leads to a more extended reionization history. This plot clearly shows that the reionization occurs in an inside-out fashion, at least when averaged over cosmologicaly relevant volumes.

A visual representation of this behaviour is shown in Figure~\ref{fig:track}. It tracks the neutral hydrogen fraction evolution around three relatively massive halos of mass \mbox{$M_\mathrm{h}=3\times 10^{12}\,\mathrm{M_\odot}$} (top panel), \mbox{$M_\mathrm{h}=6.2\times 10^{11}\,\mathrm{M_\odot}$} (middle panel) and \mbox{$M_\mathrm{h}=4.5\times 10^{11}\,\mathrm{M_\odot}$} (bottom panel) at $z=9$ (first column), $z=8$ (second column), $z=7$ (third column) and $z=6$ (fourth column). Each individual panel is a projection through a $(10\,\text{cMpc})^3$ cutout region and the quoted halo mass is at $z=5.5$, the end redshift of \thesan-1. In the early stages of reionization, the ionization front stalls close to the source because the photons from the galaxies get absorbed very quickly as the recombination time is quite short in the surrounding high density gas. The structure of these nascent \hii regions is complex with the I-front expansion happening primarily along paths with lower optical depths. Therefore, a simple picture of spherical \hii regions around sources is clearly not an accurate description of the reionization process \citep{Lee2008, Friedrich2011}. As the I-fronts reach the low density gas, they speed up, causing rapid expansion of the size of the ionized bubbles. By $z=6$, all the gas in the selected volume is ionized, except for the high density filaments and nodes. This clearly illustrates that the inside-out scenario of reionization is broadly valid. We note that there might be individual subregions where this general picture breaks down, for example, the ionizing photon budget around small halos that live in over-dense regions might be dominated by their nearby massive counterparts. A thorough investigation about the sources responsible for reionization at different gas over-densities will be taken up in a forthcoming paper. 

\begin{figure*}
	\includegraphics[width=0.99\textwidth]{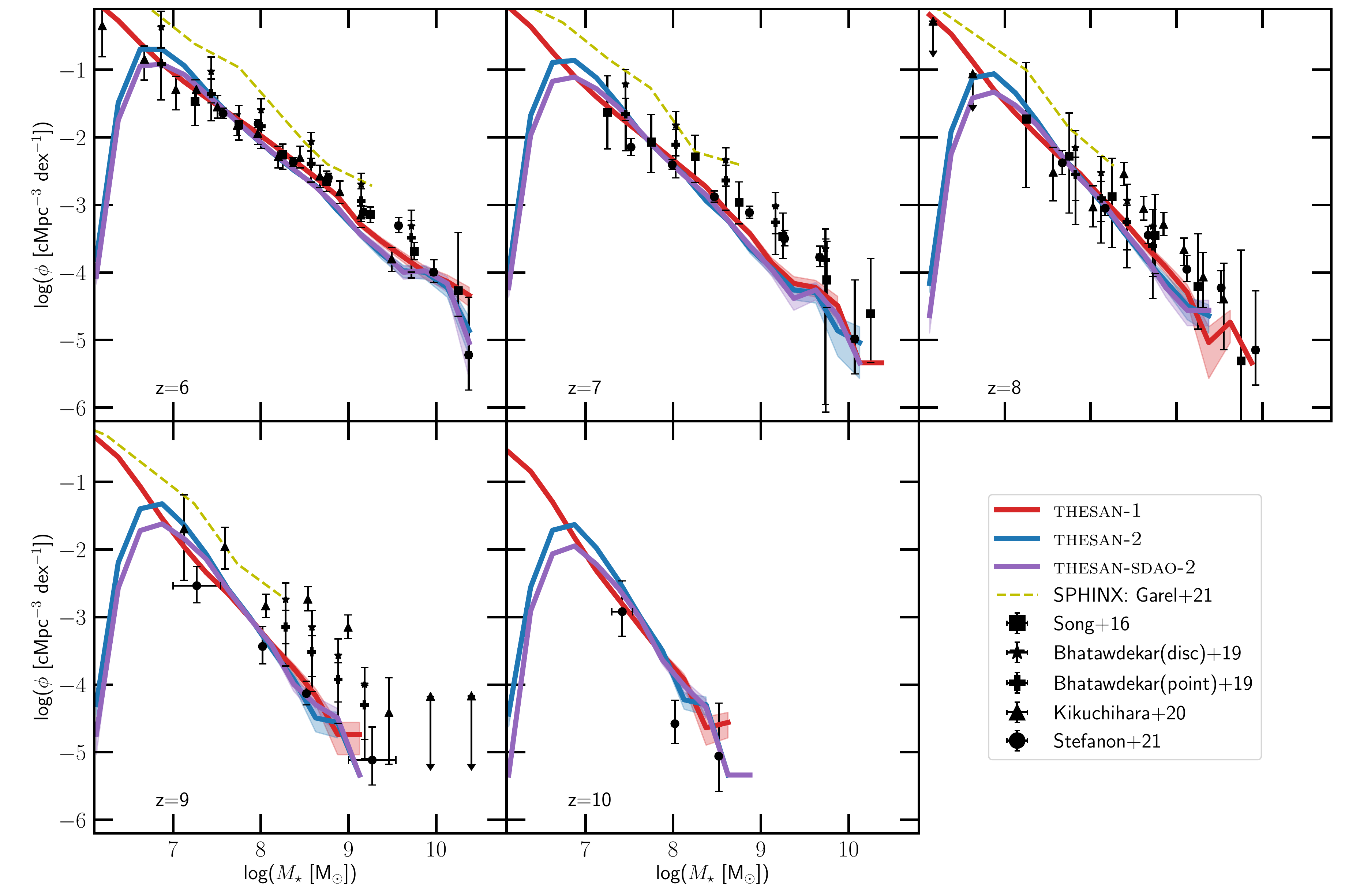}
    \caption{The galaxy stellar mass function in the \thesanone (red curve), \thesantwo (blue curve) and \thesansdao (purple curve) simulations for $z=6-10$ as indicated. The shaded regions show the Poisson noise ($\propto \sqrt{N}$, where $N$ is the number of galaxies in each bin). The observational estimates from \citet[black squares; ][]{Song2016}, \citet[black pluses for disc like source constraints and black stars for point like sources; ][]{Bhatawdekar2019}, \citet[black triangles; ][]{Kikuchihara2020} and \citet[black circles; ][]{Stefanon2021} are also shown. We also show the results from the \textsc{sphinx} simulations \citep[yellow curves; ][]{Rosdahl2018, Garel2021} for comparison.}
    \label{fig:gsmf}
\end{figure*}

\subsection{Galaxy properties}
\label{sec:galaxies}

This section focuses on the properties of the sources (galaxies and black holes) responsible for the reionization process. 

\subsubsection{Stellar properties}
\label{sec:stellar}
We first start with the stellar-to-halo-mass relation (SHMR) which quantifies the efficiency with which the DM halos convert their baryonic matter into stars \citep{Moster2010}. This relation is very important because simulations have shown that getting the right SHMR generally leads to realistic galaxy populations and therefore, the feedback models are tuned to primarily match this quantity \citep{Stinson2013, Vogelsberger2013, Kannan2014, Schaye2015}. Figure~\ref{fig:smhm} shows SHMR for galaxies in \thesanone at  $z=6$ (top left panel), $z=7$ (top middle panel), $z=8$ (top right panel), $z=9$ (bottom left panel) and $z=10$ (bottom middle panel). The solid black line shows the median and the dashed curves are the $10$ and $90$ percentiles of the simulated relation. For comparison we show the estimates from \citet{Behroozi2019} in blue and note that the simulations match this result at almost every redshift. We also show recent observational data derived using stellar mass estimates from deep IRAC/Spitzer measurements of galaxies in the reionization epoch and then adopting abundance matching to derive SHMR\footnote{The stellar mass estimates from \citet{Ceverino2017}, \citet{Tacchella2018} and \citet{Stefanon2021} have been converted to match the Chabrier IMF \citep{Chabrier2003} used in this work, by reducing them by a factor of $1.7$.}  \citep[red points; ][]{Stefanon2021}.  These points generally lie on the higher end of the simulated relation and are offset from the median by about $0.1-0.2$ dex.  Moreover, the relation seems to be steeper than both our simulated data and the \citet{Behroozi2019} relation. However, we note that the simulation results are consitent with this data within the quoted errorbars. The orange curves and the corresponding shaded region shows the SHMR estimate from \citet{Tacchella2018}, who use a simple model which assigns a SFR to each dark matter halo based on the growth rate of the halo and a redshift-independent star-formation efficiency. The slope of the relation seems to be very steep with low mass halos ($\Mh \lesssim 10^{10} \Msun$) showing similar baryon conversion efficiencies to the simulated data, but the model predicts slightly higher stellar masses in the high mass ($\Mh \sim 10^{11} \Msun$) halos.

We also show results from other recent simulation efforts to model high-redshift structure formation including \textsc{sphinx} \citep[purple dots; ][]{Rosdahl2018} and FirstLight \citep[pink dots; ][]{Ceverino2017}. \textsc{sphinx} predicts baryon conversion efficiencies that are about $0.5$ dex larger than the estimates from \citet{Behroozi2019} and \citet{Stefanon2021} and are slightly higher than the predictions from \citet{Tacchella2018}, while the estimates from FirstLight are compatible with the \citet{Tacchella2018} model. We note that these models have been specifically designed and evaluated in simulations that mainly predict observables above $z \gtrsim 6$, where the constraints are fairly limited. When some of these galaxy formation models are used to simulate galaxies down to lower redshifts, they give rise to galaxy populations that are incompatible with the observed ones. For example, when the galaxy formation model implemented in the \textsc{sphinx} simulations is used to simulate galaxies at lower redshifts (down to $z\simeq3$), it produces high stellar mass, bulge dominated galaxies with centrally peaked rotation curves \citep{Mitchell2021}. 
Finally, we also show the SHMR relation from the \textsc{croc} simulations \citep[green dashed line; ][]{Gnedin2014} which seem to underproduce stellar masses of the most massive halos ($\Mh \sim 10^{11} \Msun$), which might point to the fact that the stellar feedback model over-suppresses star formation in high-mass objects \citep{Zhu2020}. 

\begin{figure}
	\includegraphics[width=0.99\columnwidth]{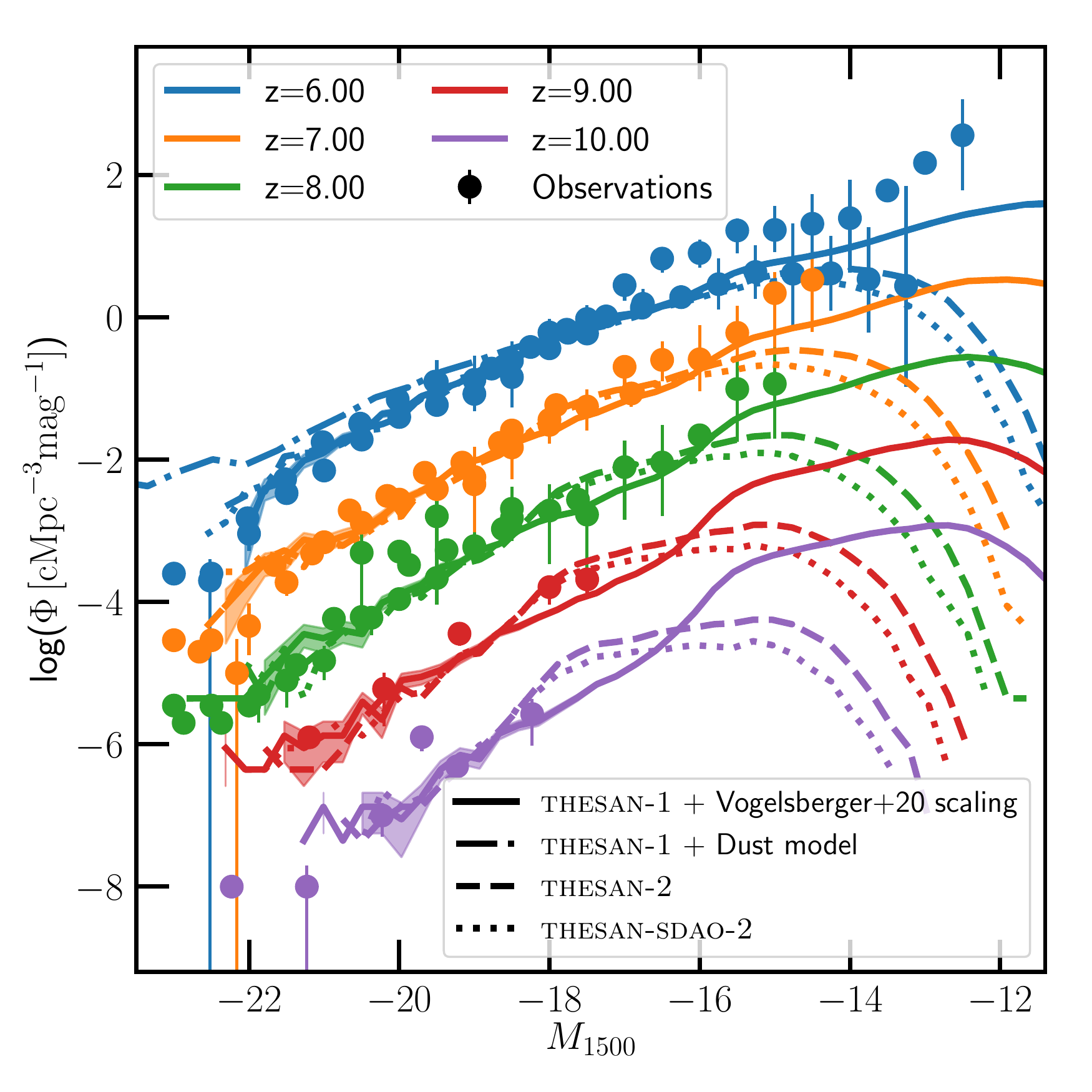}
    \caption{UV luminosity functions at $z=6-10$, for the \thesanone (solid curves), \thesantwo (dashed curves) and \thesansdao (dotted curves) simulations. An empirical relation presented in \citet{Gnedin2014} is used to account for dust attenuation, with the opacity scaled according to the redshift dependent dust-to-metal (DTM) ratio given in \citet{Vogelsberger2020}. We also show (for $z=6$ and \thesanone) the dust correction obtained using the dust mass estimates from the empirical dust model used in \thesan (blue dot-dashed curve). The observational estimates are taken from \citet{Bouwens2015, Bouwens2017}, \citet{Finkelstein2015}, \citet{McLeod2016}, \citet{Livermore2017}, \citet{Ishigaki2018} and \citet{Atek2018}. We note that the luminosity functions are offset by $\Delta \mathrm{log}(\Phi) = -(z-8)$. The simulated UV luminosity function matches the observational estimates over a wide range of magnitudes. The superior resolution of \thesanone allows it to model the very faint galaxies (up to \mbox{$M_{1500} \sim -12$}), while \thesantwo and \thesansdao show a turn over at about \mbox{$M_{1500} \sim -15$} due to resolution effects and lack of small scale power.
    }
    \label{fig:uvlf}
\end{figure}

The SHMR relies on accurately populating the simulated DM halos with the observed stellar masses of galaxies. We therefore directly compare the simulated stellar mass function (SMF) to the observationally inferred one in Figure~\ref{fig:gsmf}, which shows the galaxy stellar mass function for \thesanone (red curves) at $z=6-10$ as indicated. The plot also shows the results from the \thesantwo (blue curves) and \thesansdao (purple curves) simulations. Comparing the stellar mass functions for \thesanone and \thesantwo, we conclude that the results are well converged above {$M_\star \gtrsim 10^7 \Msun$}. Below this limit, the \thesantwo SMF falls off due to resolution effects. Similarly, the low mass cutoff in the matter power spectrum in \thesansdao causes the SMF to fall off at a slightly higher stellar mass.  

The observational estimates from \citet[black squares; ][]{Song2016}, \citet[black pluses for disc like source constraints and black stars for point like sources; ][]{Bhatawdekar2019}, \citet[black triangles; ][]{Kikuchihara2020} and \citet[black circles; ][]{Stefanon2021} are shown for comparison. At low redshifts ($z<8$), the simulations generally match the observational results below {$M_\star \lesssim 10^9 \Msun$} but are lower by less than $\sim 0.1$ dex for the higher mass galaxies. This reflects the different slopes of the SHMR in simulations and observations (see Figure~\ref{fig:smhm}). At higher redshifts the simulations match the observational estimates from \citet{Stefanon2021} throughout the entire range, but are lower than the estimates from \citet{Bhatawdekar2019} and \citet{Kikuchihara2020} at the high mass end. Finally we also show the SMF for the \textsc{sphinx} simulations \citep[dashed yellow curves; ][]{Garel2021} which generally lies about $0.3$--$0.5$ dex above the \thesan results reflecting the higher star formation efficiency implied by the SHMR. At $z\lesssim8$, many of the observationally inferred estimates seem to favour the \thesan results, except for the \citet{Bhatawdekar2019} estimates based on disc-like source profiles. If point-like source profiles are assumed then the SMF estimates are in better agreement with our results. These inconsistencies in the observations mainly arise from uncertainties in the lensing models used in these works. Upcoming observations with the \jwst will help to resolve the tension between different observations and place much better constraints on the stellar populations of high-$z$ galaxies.

 \begin{figure}
	\includegraphics[width=0.99\columnwidth]{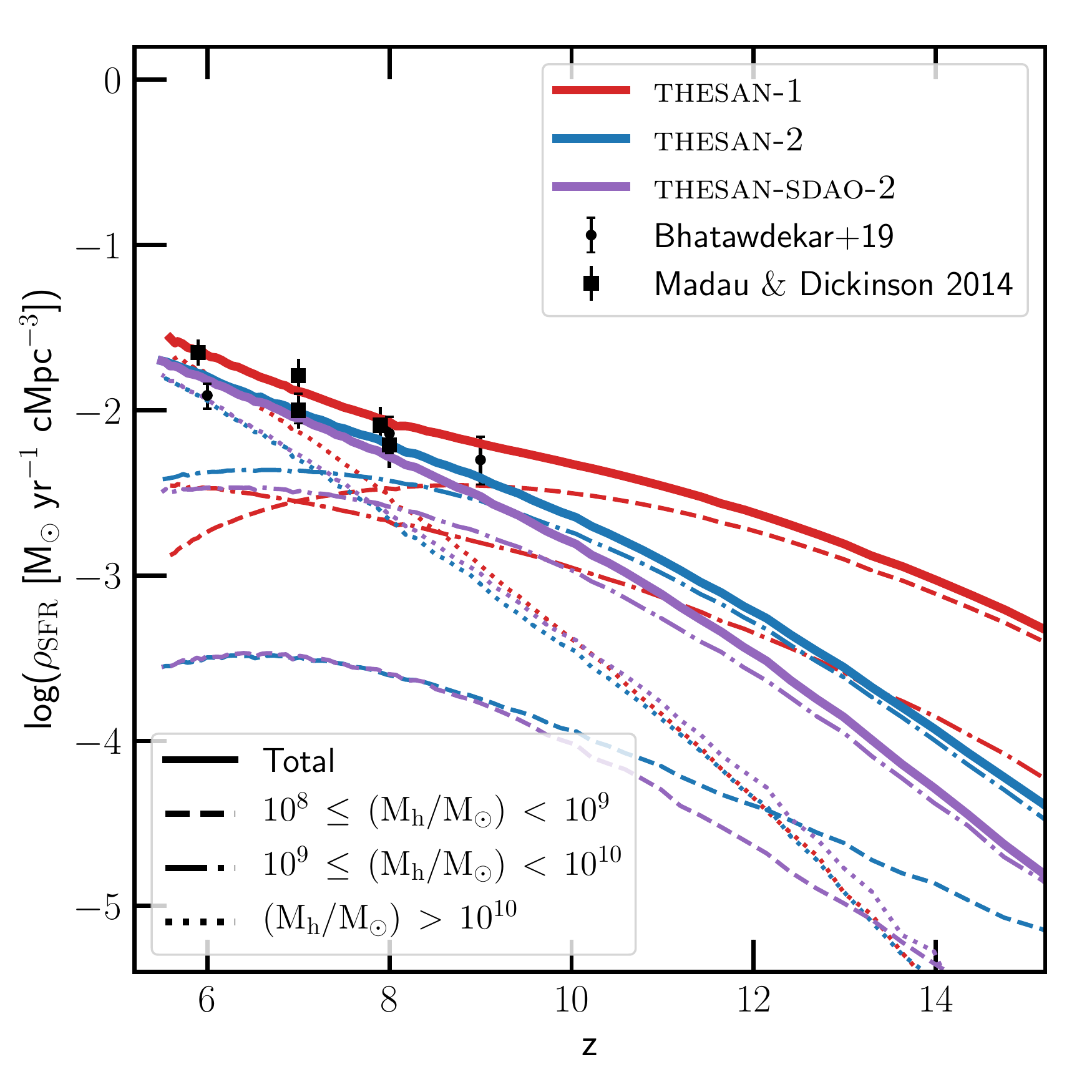}
    \caption{Evolution of the star formation rate density. Contribution from low, medium and high mass halos to the total star formation rate are also shown as dashed, dot-dashed and dotted curves, respectively. Observational estimates from \citet{Madau2014} and \citet{Bhatawdekar2019} are shown in black squares and circles, respectively. While the SFRD below $z=8$ is well converged, the low resolution simulations show an order of magnitude lower SFR rate at high redshifts, due to their inability to resolve the low mass halos.}
    \label{fig:sfrd}
\end{figure}

Figure~\ref{fig:uvlf} shows the UV luminosity functions in the \thesanone (solid curves), \thesantwo (dashed curves) and \thesansdao (dotted curves) simulations at $z-6-10$, as indicated. This is obtained by summing up the radiation output at rest frame \mbox{$1500$ \AA} (using BPASS tables) of all the stars in the identified subhalo. We note that due to the probabilistic nature of the star formation routine, the star formation history will only be sparsely sampled in halos with low SFR. These halos will have long periods with zero star formation interspersed with sudden jumps in SFR as a new particle is stochastically spawned (the mass of the newly formed star particle is similar to the baryonic mass resolution of the simulation, which for \thesanone is \mbox{$5.82 \times 10^5$\,M$_\odot$}). This young and massive star will then dominate the whole radiation output of the galaxy especially if the mass of the galaxy is very close to the resolution limit of the simulation. This will adversely affect the UV luminosity function of the simulation. In order to overcome this numerical artifact, the age and mass of stars formed less than \mbox{$5$ Myr} ago are smoothed over a timescale given by $t_\mathrm{smooth} = \Sigma M_\star (< 5 \mathrm{Myr}) / \mathrm{SFR_{gal}}$, where $\mathrm{SFR_{gal}}$ is the instantaneous SFR of that particular galaxy calculated by summing up all the SF probabilities of the cells in the EoS \citep[see][for more details]{Springel2003}. This smoothing procedure is only done for halos with \mbox{$t_\mathrm{smooth}> 5$ Myr}. We note that this only affects halos close to the resolution limit and allows for a more faithful prediction of the simulated UV luminosity function. The high mass end of the UV luminosity function is very sensitive to attenuation by dust. An empirical relation presented in \citet{Gnedin2014} is used to account for dust attenuation and the opacity is scaled according to the redshift dependent dust-to-metal (DTM) ratio given in \citet{Vogelsberger2020}.  While, the \thesan simulations include a semi-empirical model to estimate the dust masses in galaxies, the dust-to-mass and dust-to-metal ratios are lower than what is expected (see Section~\ref{sec:metalanddust} for more details). These low dust masses lead to dust attenuation in luminous halos being not strong enough to match the observational estimates (dot-dashed blue curve). The scaling relation derived in \citet[red curve; ][]{Vogelsberger2020} on the other hand does a good job in matching the observations. The observational estimates are shown as coloured circles.

While the high mass end of this relation is pretty robust, the estimates beyond $M_\mathrm{{UV}}\sim-15$ have quite a large scatter. These low luminosity galaxies are difficult to observe at such high redshifts and therefore the only estimates come from lensed galaxies observed in the Hubble Frontier Fields.  There are significant differences between published results because of different lensing magnification models and their associated complex uncertainties. For example,  \citet{Bouwens2017} derive a shallow faint end slope of $\alpha \sim -1.9$, while \citet{Livermore2017} and \citet{Ishigaki2018} estimate a much steeper slope of about $\sim -2.1$. \citet{Atek2018} showed that the size distribution  and the choice of lens model leads to large differences at magnitudes fainter than $M_\mathrm{{UV}} = -15$, where the magnification factor becomes highly uncertain. They favour a model where the luminosity function turns over at these faint magnitudes. More importantly, they conclude that robust constraints on the UVLF at faint magnitudes remain unrealistic with current observational techniques. Future robust measurements with \jwst are needed to resolve this discrepancy.

Constraining the low mass slope of the UVLF is very important for figuring out the mechanism of the reionization process. Models that rely on low mass halos to contribute the bulk of the ionizing photon budget indicate that full reionization by $z \sim 6$ can only be achieved if the observed luminosity function is extrapolated by two orders of magnitude below ($\mathrm{M_{UV}} \sim -13$) the current observational limit of the \hst \citep{Bouwens2015, Finkelstein2015, Finkelstein2019}. We note that \thesanone, with its superior resolution, predicts a relatively shallow faint end slope. The UVLF only starts to turn over beyond $M_\mathrm{{UV}} \sim -13$  due to resolution effects. This matches the estimates from \citet{Bouwens2017}. \thesantwo, on the other hand, shows a turnover at magnitudes fainter than $M_\mathrm{{UV}} \sim -15$ and is more in line with the estimates from \citet{Atek2018}. This is purely due to the lower resolution of the \thesantwo simulation. In a similar vein, the cutoff in the matter power spectrum at small scales in the \thesansdao model also leads to a turnover at about the same magnitude as \thesantwo. We note that we have not tuned these different simulations to match the different observation results. The fact that they turn over at the same point as some of the observations is purely serendipitous. However, these simulations can be used to test the different reionization models that assume these different low mass slopes for the UVLF.

   \begin{figure}
	\includegraphics[width=0.99\columnwidth]{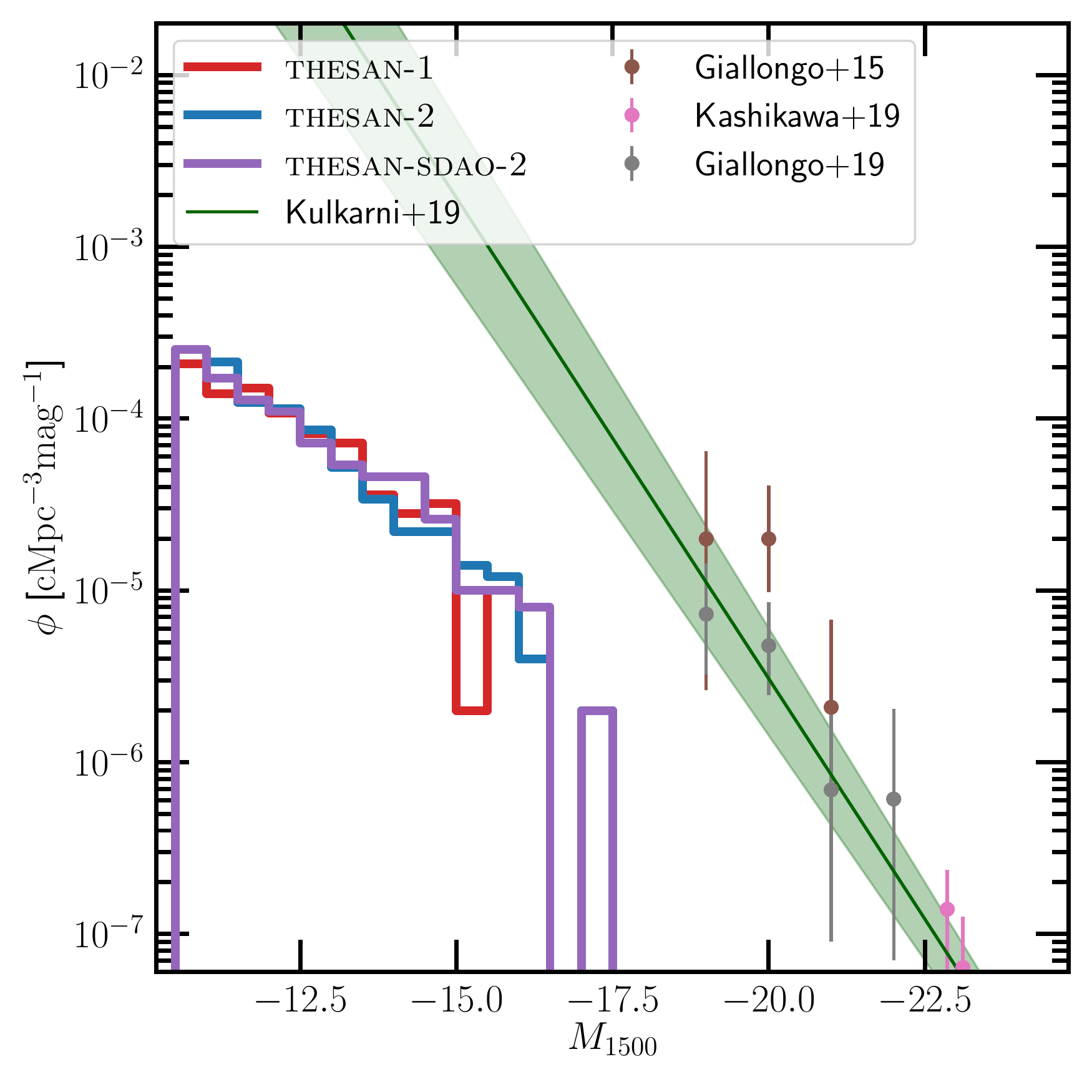}
    \caption{The quasar luminosity function at $z=6$, for the \thesanone (red curve), \thesantwo (blue curve) and \thesansdao (purple curve) simulations. Observational estimates from \citet{Kashikawa2015} and \citet{Gianllongo2015, Gianllongo2019} are show as coloured circles. We also show the derived AGN UV luminosity function from \citet[green shaded region; ][]{Kulkarni2019_QLF}. The simulation volume is not large enough to contain the highly luminous but scarcely abundant massive black holes observed at high redshift.}
    \label{fig:qlf}
\end{figure}

Figure~\ref{fig:sfrd} shows the evolution of the star formation rate density (SFRD) in the \thesanone (red curves), \thesantwo (blue curves) and \thesansdao (purple curves) simulations. For comparison, the plot also shows the observationally inferred estimates from \citet{Madau2014} and \citet{Bhatawdekar2019}, plotted as black squares and circles, respectively. The differences between the observational estimates likely arise due to the fact that these works integrate the SF down to different UV magnitude limits ($M_\mathrm{UV} \sim -17.5$ for \citealt{Madau2014} and $M_\mathrm{UV} \sim -13.5$ for \citealt{Bhatawdekar2019}). The simulations on the other hand include all the SF in the simulation volume, down to the resolution limit.   While all three simulations show roughly similar SFRD at low redshift, they diverge considerably at $z\gtrsim8$. This behaviour can be better understood by splitting the total SFRD into contributions from low mass (dashed curves; $10^8\,\Msun < \Mh < 10^9\,\Msun$), medium mass (dot dashed curves; $10^9\,\Msun < \Mh < 10^{10}\,\Msun$) and high mass (dotted curves; $\Mh > 10^{10}\,\Msun$) halos for each simulation. The star formation rate in high mass halos is reasonably converged with all three simulations showing similar values. However, the contribution of these halos to the total SFRD is negligible at high redshifts and they only start to dominate below $z\lesssim8$ at which point they all show similar total SFRD. The star formation in low mass halos, however, shows a marked difference between the three runs. Above $z>8$, almost all the star formation in \thesanone occurs in these low mass halos, while they are subdominant in the other two runs. This is because \thesantwo is unable to resolve these low mass halos and hence their contribution to the total star formation rate is diminished. Similarly, the \thesansdao model's ability to suppress low mass galaxy formation also results in a very low SF in these halos. In these two simulations, the medium mass halos dominate the total SFRD at high redshifts. We also note that all three simulations show a dip in the SFR, especially in low mass halos as reionization progresses. This is likely due to  photoheating feedback by the reionization process which reduces the  accretion of gas onto these small halos \citep{Gnedin2014, D18, Wu2019a}. We plan to investigate this phenomenon further in a forthcoming paper.

\subsubsection{Black hole properties}
\label{sec:BH}
Figure~\ref{fig:qlf} focuses on the properties of black holes by plotting the quasar luminosity function (QLF) in the three simulations as indicated, with the estimates from observations \citep{Kashikawa2015, Gianllongo2015, Gianllongo2019} plotted as coloured circles and the derived AGN UV luminosity relation from \citet{Kulkarni2019_QLF} shown as the green shaded region. Almost all the black holes in our model are in the high accretion phase of their evolution (the low accretion regime only starts to become important below $z\sim1.5$; \citealt{Kannan2017}). In this `quasar' phase, we assume that the bolometric luminosity is proportional to the mass accretion rate with a radiative conversion efficiency of $0.2$ \citep{Weinberger2018}. The UV luminosity is then calculated assuming the SED parametrization outlined in \citet{Lusso2015}. We note that since the black holes primarily reside in high mass halos, whose properties are similar in all the three simulations considered here, the QLF is well converged. The relatively low simulation volume restricts our ability to simulate the really high luminosity AGNs that have been recently observed  at these redshifts. The number of low luminosity AGN also seems to be lower than the extrapolated estimates of \citet{Kulkarni2019_QLF}. These low luminosity AGN, therefore, do not contribute significantly ($<1\%$ of the global ionizing photon budget) to the reionization process in our simulations.

\subsubsection{Metal and dust enrichment}
\label{sec:metalanddust}

\begin{figure}
	\includegraphics[width=0.99\columnwidth]{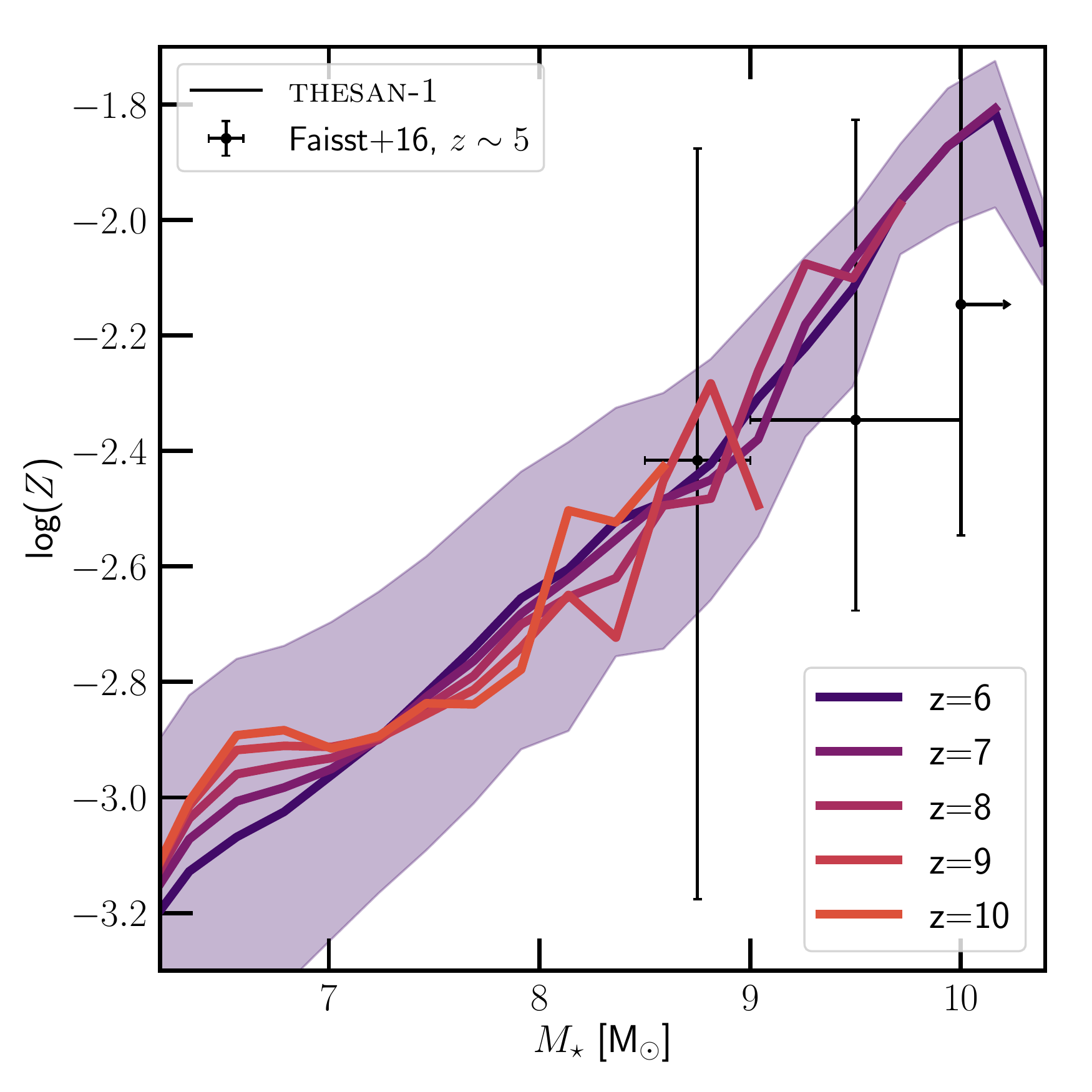}
    \caption{The mass-metllicity relation of gas in the galaxies present in the \thesanone simulation at $z=6-10$. The shaded region shows the typical dispersion in this relation. The observational results at $z~\sim5$ are shown as black circles \citep{Faisst2016}. The simulations do a good job of reproducing the mass-metallicity relation of high redshift galaxies.}
    \label{fig:mmr}
\end{figure}

The metal and dust contents of galaxies play an important role in the galaxy evolution process. They account for a large fraction of the radiative cooling rate which in turn controls the gas accretion and star formation rates.  They also regulate the  attenuation and escape of radiation from the host galaxy \citep{Benson2003}. It is therefore important for simulations to properly model these quantities in order to get accurate galaxy properties. We first start with the gas phase metallicity-stellar mass relation at $z=6-10$ predicted by the \thesanone simulation (Figure~\ref{fig:mmr}). The shaded region shows the typical dispersion in this relation. The simulation predicts that there is no evolution in this relation with redshift. The galaxies start small with low metallicities, beginning at the lower left corner of the phase-space and then move along the relation as they grow larger. This is consistent with the very weak evolution of the normalisation of the mass-metallicity  relation outlined in \citet{Torrey2019}.  Estimating the metallicities is a difficult endeavour for high-redshift galaxies. Therefore, there are only a small number of observations with which we can compare our simulation results. The black circles show the results from \citet{Faisst2016} who use a calibrated relation between the depth of four prominent rest-UV absorption complexes and metallicity of local galaxies and extrapolate it to $z\sim 5$. The simulations are in broad agreement with the observational results.

\begin{figure*}
	\includegraphics[width=1.99\columnwidth]{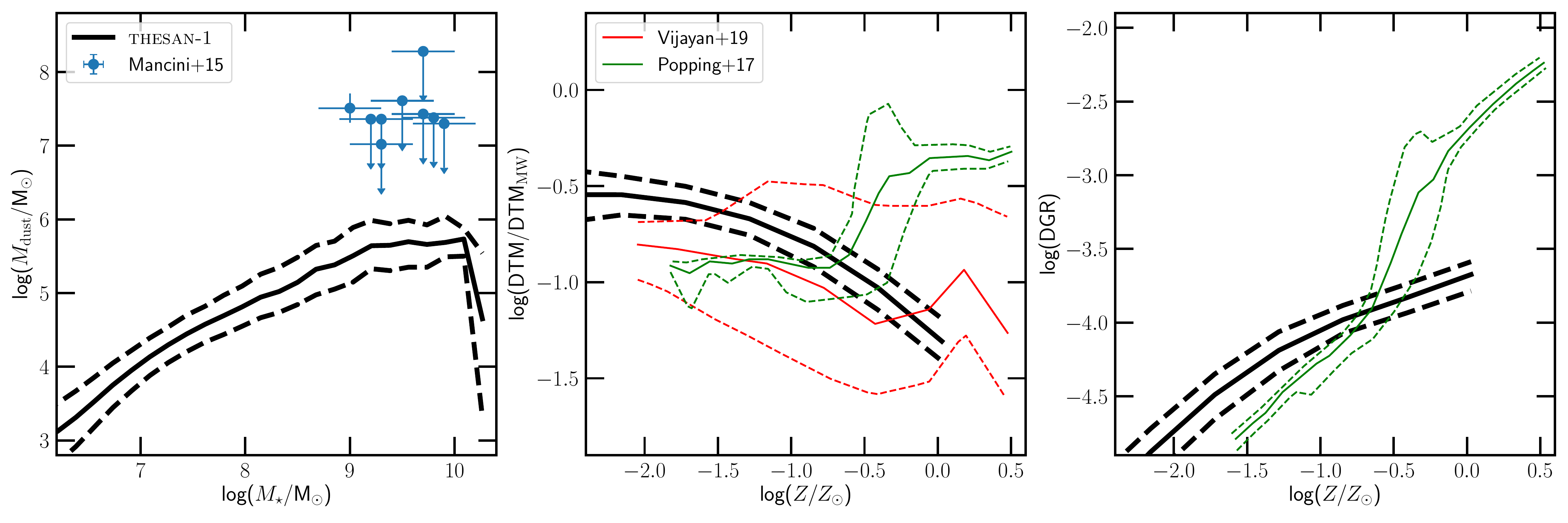}
    \caption{The properties of dust at $z=6$ in the \thesanone simulation. The left panel shows the total dust mass of the galaxy as a function of its stellar mass compared with observations from \citet{Mancini2015}. The middle panel plots the galaxy averaged dust-to-metal (DTM) ratio as function of the metallicity of the galaxy, compared with estimates from semi-analytic models of \citet{Popping2017} and \citet{Vijayan2019}. The right panel shows the dust-to-gas (DGR) ratio as a function of metallicity. The predicted dust content in high metallicity environments is low when compared to observational estimates and other semi-analytic works.}
    \label{fig:dust}
\end{figure*}

Figure~\ref{fig:dust} shows the various properties of dust in galaxies at $z=6$ in \thesanone as predicted by the empirical dust model described in Section~\ref{sec:met:dust}. We start with the left panel which shows the dust mass within two times the half mass radius of stars as a function of the stellar mass (calculated within the same radius) of the galaxy. The observational estimates from \citet{Mancini2015} are shown as blue circles. We immediately see that the simulation predicts about an order of magnitude lower dust masses compared to the  observations. However, we note that most of the observational estimates are upper limits and therefore our results are still compatible with the data. The middle panel shows the dust-to-metal ratio (DTM) normalised to the Milky-Way (MW; 0.44) value as a function of the gas-phase metallicity of the galaxy. Since the observational estimates of this quantity are practically non-existent we compare our results to two recent semi-analytic models of dust formation. Our simulations predict a monotonically decreasing DTM with metallicity. At low metallicities (one hundredth solar) the DTM is relatively high with a value of  $\sim 0.14$ and then decreases to a value of $\sim 0.02$ at solar metallicity.  The \citet{Popping2017} model on the other hand shows a flat relation with a DTM value of $\sim 0.07$ up to about a third solar and then dramatically increases in value and saturates at about \mbox{$\mathrm{DTM}\sim0.3$}. This behaviour is attributed to a combination of low condensation efficiencies and the rapid growth of dust in the ISM at high metallicities. On the other hand, the model by \citet{Vijayan2019} shows a more shallow relationship, with the DTM values showing a slight decrease with increasing metallicity. This is because they explicitly model the dust content in molecular clouds and the intercloud medium. Our simulation results agree well with the \citet{Vijayan2019} semi-analytic model at low metallicities but produce relatively low dust masses at the higher metallicity end. The right panel shows the dust-to-gas ratio (DGR) as a function of metallicity compared to the estimates from the \citet{Popping2017} model. Reflecting the behaviour in the DTM relation, the simulations show a smooth increase in DGR as the metallicity increases, while the semi-analytic model shows a sharp increase at the same metallicity value where it predicts an increase in the DTM relation.

 \begin{figure}
	\includegraphics[width=0.99\columnwidth]{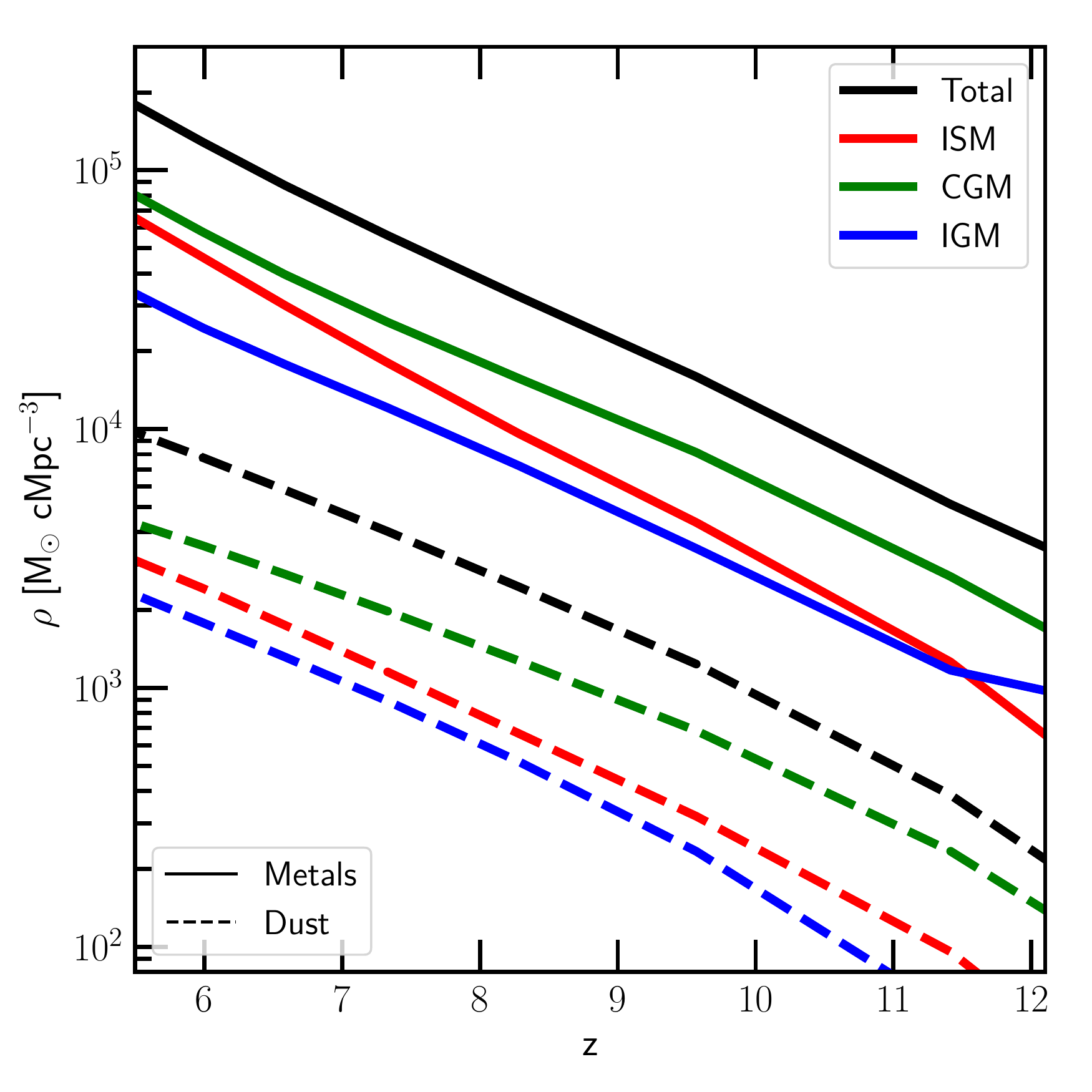}
    \caption{Evolution of the comoving mass density of metals (solid curves) and dust (dashed curves) in the ISM (red curves), CGM (green curves) and IGM (blue curves) of the \thesanone simulation. Feedback driven winds efficiently transport the metals and dust out of the ISM of the galaxy, helping to enrich the CGM and IGM.}
    \label{fig:metaldust}
\end{figure}

\begin{figure*}
	\includegraphics[width=0.99\textwidth]{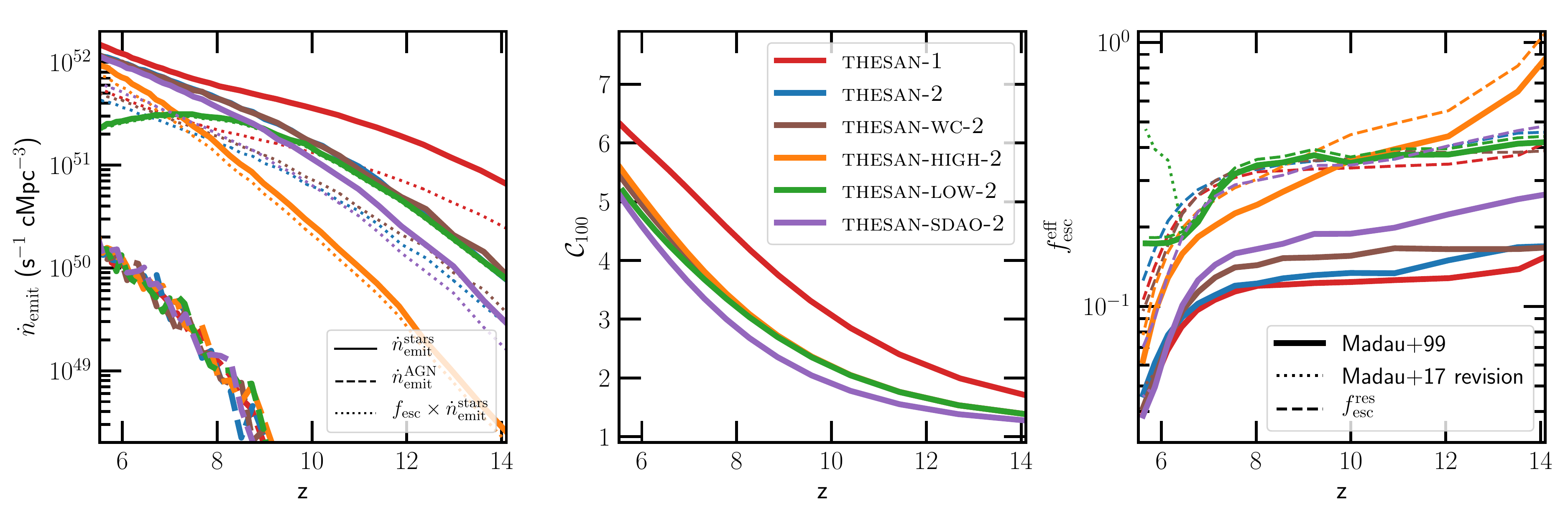}
    \caption{Left panel: The total ionizing emissivity of stars (solid curves) and black holes (dashed curves) as a function of redshift. Middle panel: The clumping factor for gas with over-densities less than $100$. Right panel: The effective escape fraction calculated using the equations outlined in \citet{Madau1999} and \citet{Madau2017}. The effective escape fraction of ionizing photons is about $15-20\%$ in our fiducial model, which is in agreement with previous estimates \citep{Gnedin2016, Rosdahl2018}.}
    \label{fig:fesc}
\end{figure*}

Finally, we note that the amount of dust predicted by the simulations, at least in high mass halos, is too low compared to the scaling relations obtained in \citet{Vogelsberger2020}. They required a {$\mathrm{DTM}=0.11$} at $z=6$ in order to match the high mass slope of the UVLF. In Figure~\ref{fig:uvlf} we have shown that using the DTM predicted from the simulations does not attenuate the light from high mass galaxies enough to match the observed UVLF, and the much higher values derived in \citet{Vogelsberger2020} are necessary  in order to comply with the observations. From these plots we can probably conclude that the empirical dust model as used in our simulations does not produce enough dust in galaxies to match the available observational constraints.

Up until now,  we have only discussed the dust and metal content in the ISM of galaxies. Figure~\ref{fig:metaldust} aims to quantify the dust and metal content in three different phases; ISM, CGM and IGM. The ISM is defined as all gas within twice the stellar half mass radius, the CGM is all gas within the identified FoF groups but not part of the ISM and the amount of metal and dust in the IGM is the total metal/dust mass in the simulation minus the contributions from the ISM and the CGM. The plot shows the comoving mass density of metals and dust as a function of redshift in \thesanone. All the metals and part of the dust is formed in the ISM of galaxies via stellar evolutionary processes \citep{Hoyle1946, Dwek1998}. The rest of the dust is produced by metal depletion onto existing dust grains \citep{Draine1990}. This process also mainly takes place in the high denity ISM gas. Large scale galactic winds generated by stellar and black-hole feedback then transport the metals and dust out into the CGM and even the IGM \citep{Vogelsberger2019, Kannan2021}. We find that  most of the dust and metals reside in the CGM of galaxies. This is true for all the redshifts considered, in agreement with low redshift results from \citet{Torrey2019}. The ISM is only the second biggest reservoir of metals and dust in the Universe. There is also a significant enrichment of the IGM, implying that galactic winds are efficient in transporting material out to very large distances. Recent observations have detected this large scale metal distribution around galaxies \citep{Bosman2017} which suggests that the CGM/IGM was already significantly enriched with metals within the first billion years of the Universe.

\subsection{Effective escape fractions}

The fluid description of the radiation field does not allow us to track individual photon paths and hence it is difficult to estimate escape fractions directly from the simulations. Ray-tracing codes can be used in post-processing to determine this quantity \citep{Rosdahl2018} and we are currently pursuing this approach in a forthcoming paper. 
However, in this work we use a simple,  zeroth order equation first outlined in \citet{Madau1999} to estimate the effective fraction of ionizing photons that escape in to the IGM. It accounts for the production and absorption of ionizing LyC radiation in a clumpy and expanding medium and is given by
\begin{equation}
    \frac{\mathrm{d}Q_\mathrm{\ion{H}{II}}}{\mathrm{d}t} = \frac{\dot{n}_\mathrm{ion}}{{\langle n_\mathrm{H}}\rangle} - \frac{Q_\mathrm{\ion{H}{II}}}{\bar{t}_\mathrm{rec}} \, .
    \label{eq:madau}
\end{equation}
$Q_\mathrm{\ion{H}{II}}$ and $\langle n_\mathrm{H}\rangle$ denote the mass weighted ionized hydrogen fraction and the physical mean hydrogen density respectively. $\dot{n}_\mathrm{ion}$ is the total ionizing photon rate and is given by \mbox{$\dot{n}_\mathrm{ion}=f_\mathrm{esc}^\mathrm{eff}\dot{n}_\mathrm{emit}$} where $f_\mathrm{esc}^\mathrm{eff}$ is the effective escape fraction  and $\dot{n}_\mathrm{emit}$ is the total number of ionizing photons emitted by the radiation sources present in the simulation. $\bar{t}_\mathrm{rec}$ is the average recombination rate which is defined as 
   $\bar{t}_\mathrm{rec} = {1}/{\langle n_\mathrm{H}\rangle \alpha \mathcal{C} }$,
where $\alpha$ is the hydrogen  case B recombination coefficient at a temperature of $\sim 10^4$~K and $\mathcal{C}$ is the clumping factor. Following other recent works we set $\mathcal{C} = \mathcal{C}_{100}$ which is the clumping factor ($\langle\rho^2\rangle/\langle\rho\rangle^2$) calculated for gas with over-densities less than $100$.

The left panel of Figure~\ref{fig:fesc} shows the emissivity ($\dot{n}_\mathrm{emit}$) of stars (solid curves) and black-holes (dashed curves) in the simulations as indicated. The dotted curve gives $\dot{n}_\mathrm{ion}$ after the unresolved escape fraction ($f_\mathrm{esc}$) has been taken into account. It is immediately clear that the AGN contribution to the total ionizing photon budget is minimal. For the \thesanhigh and \thesanlow simulations we assume that all halos above/below $10^{10}~\Msun$  do not contribute to the photon budget (respectively). It is quite interesting to note that almost all simulations show a monotonically increasing $\dot{n}_\mathrm{emit}$, while the drop in SFR in low mass halos due to the reionization process, causes $\dot{n}_\mathrm{emit}$ in \thesanlow to drop after attaining a peak at about $z\sim7$. Recent works have favoured a drop in the emissivity below $z=6$, in order to explain the observed \lyalpha opacitites after overlap \citep{Keating2020, Ocvirk2021}, although we note that this can also be explained by a drop in the escape fraction from galaxies, as they become more massive.

The middle panel (of Figure~\ref{fig:fesc}) shows $\mathcal{C}_{100}$ as a function of redshift for the different simulations. The clumping factors start low with a value of about $\sim 1.5$ at $z=14$ and increase to about $6$ at $z=5.5$. The higher resolution \thesanone simulation shows a slightly higher clumping factor ($\sim 15\%$ larger) than the other medium resolution runs. We note that the this is similar to the values obtained in other reionization simulations \citep{Kaurov2015}. The solid curves in the right panel (of Figure~\ref{fig:fesc}) show the the effective escape fractions ($f_\mathrm{esc}^\mathrm{eff}$) obtained by solving Eq.~\ref{eq:madau}. The two fiducial simulations have effective escape fractions (at $z>7$) of about $10-20\%$, in agreement with previous simulations and analytic works \citep{Gnedin2016, Madau2017}. \thesanwc shows a slightly larger effective escape fraction, because the input unresolved escape fraction is higher. The other three simulations require  higher escape rates in order to match the observed neutral fraction evolution. This is because, by construction, only the high and low mass halos contribute photons in the \thesanhigh and \thesanlow simulations. Similarly, the lack of low mass halos in \thesansdao reduces the total SFR which in turn reduces the total number of photon emitted (see left panel of Figure~\ref{fig:fesc}).  As $\dot{n}_\mathrm{emit}$ decreases, $f_\mathrm{esc}^\mathrm{eff}$ has to increase correspondingly in order to get a similar ionization history. However, we note that the effective escape fraction is a combination of both resolved and unresolved escape fractions \mbox{($f_\mathrm{esc}^\mathrm{eff} = f_\mathrm{esc} f_\mathrm{esc}^\mathrm{res}$)}. When we factor out the differences caused by the distinct $f_\mathrm{esc}$ factors, then $f_\mathrm{esc}^\mathrm{res}$ has very similar values across all simulations (dashed curves). This implies that the resolved escape fraction is constant throughout and does not depend on galaxy mass and has a very weak dependence on redshift. However, we need to take this result with a grain of salt because there are a lot of assumptions that enter this calculation. A more thorough investigation using ray-tracing RT codes is needed to more reliably model the escape fraction and related physics from galaxies \citep{Smith2019,Smith2020}. We do, however, note that some inaccuracies in the escape fraction estimates might arise due to the fact that the feedback model temporarily decouples winds from the hydrodynamics, which reduces the ability to create low density channels through which radiation can escape \citep{Trebitsch2017}. Moreover, the effective equation of state ISM model gives rise to a smooth gas distribution that will likely result in too high volume-weighted densities, which may artificially reduce the escape fractions.

\begin{figure*}
	\includegraphics[width=0.495\textwidth]{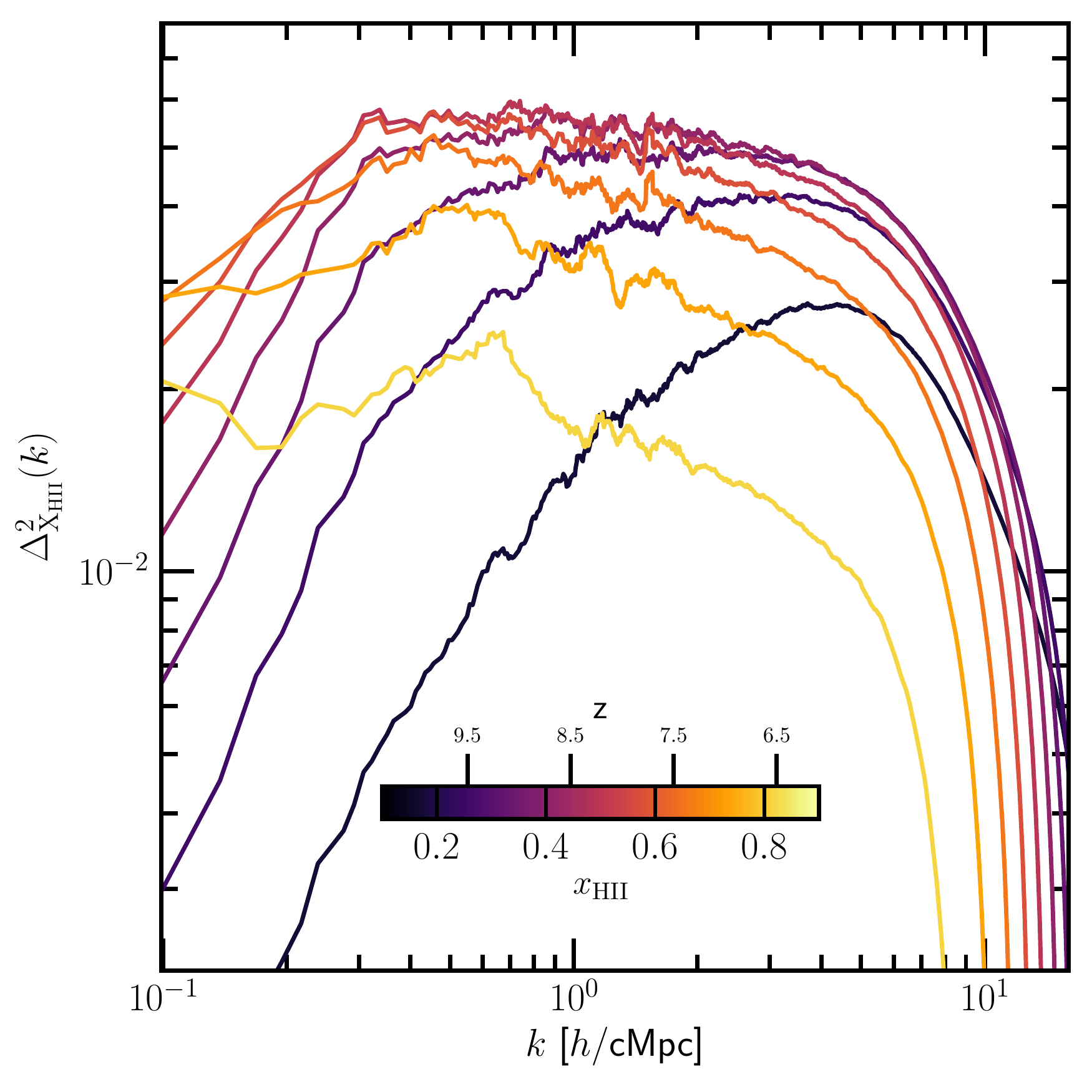}
	\includegraphics[width=0.495\textwidth]{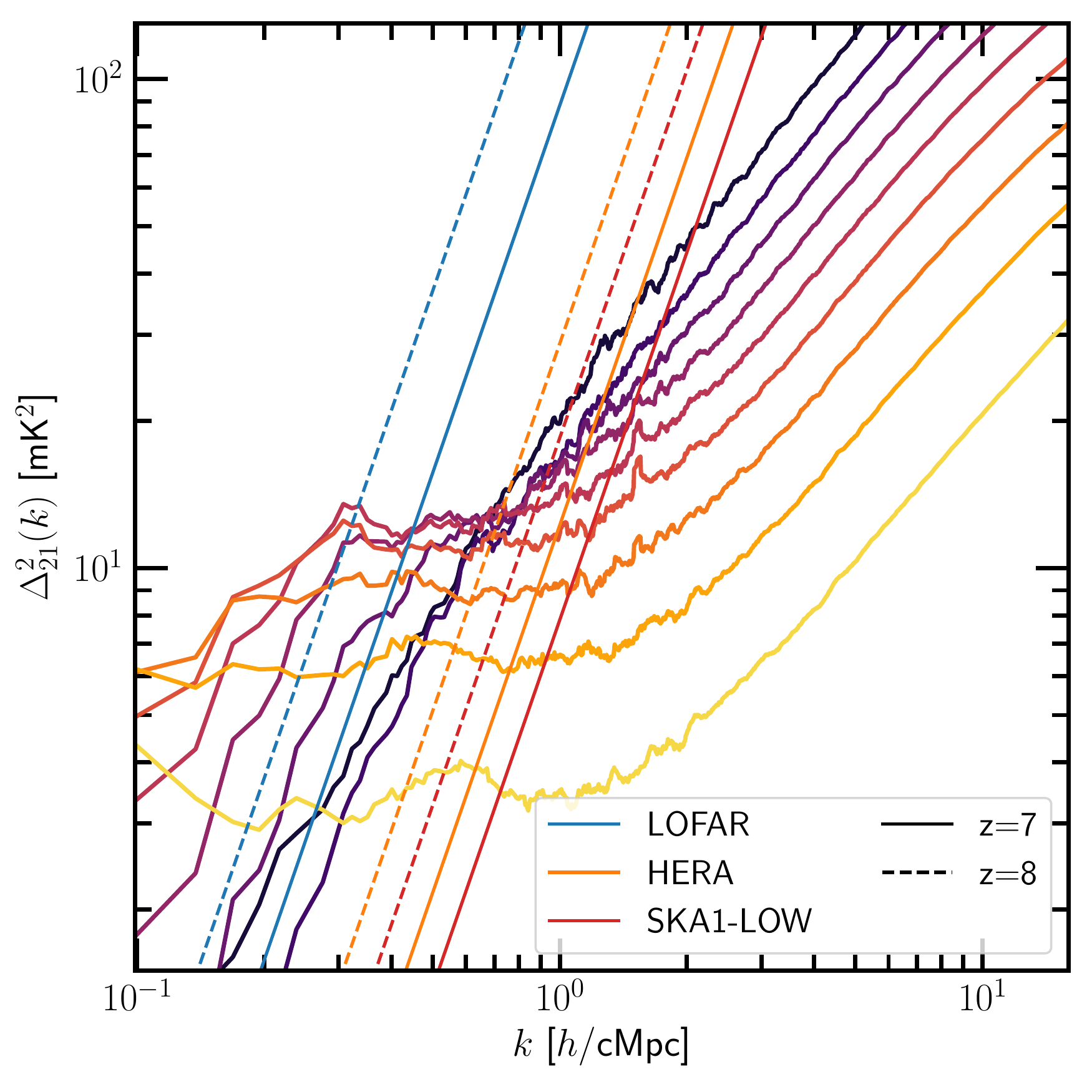}
    \caption{The left panel shows the power spectrum of ionized hydrogen in \thesanone at different volume weighted ionized fractions ranging from $0.1-0.9$, in increments of $0.1$. The right panel shows the same for the redshifted 21\,cm signal. The bubbles start out small and get bigger and merge with each other. This is reflected in the fact that the power spectrum peaks at large-$k$ for low ionization fractions and then shifts to smaller and smaller wavenumbers as the reionization process progresses. These additional fluctuations are also imprinted in the 21\,cm signal which can be measured by current and upcoming instruments like \lofar, \hera and \ska, whose sensitivity limits at $z=7$ and $z=8$ \citep{Pober2014, Kulkarni2016} are also shown as indicated. }

    \label{fig:Pk}
\end{figure*}

The steep decrease in the effective escape fraction at $z<7$ is likely due to the inability of this simple model to asses the impact of Lyman Limit Systems (LLS).  These are high-density regions that occupy a small portion of the volume but are able to keep a significant fraction of their hydrogen in neutral form. At very low neutral fractions, most of the photons are absorbed by these systems and only a small fraction is available for IGM ionization. Another possible explanation is that, as the mean free path increases, there is a delay between the photons being emitted and absorbed. This implies that $\dot{n}_\mathrm{emit}$ from an earlier time needs to enter Eq.~\ref{eq:madau}. Quantifying this delay time is beyond the scope of this paper (however see \citealt{Gnedin2016} for an empirical relation). \citet{Madau2017} suggests a modification to Eq.~\ref{eq:madau} that takes into account the impact of LLS. It replaces $\langle n_\mathrm{H}\rangle$ with \mbox{$\langle n_\mathrm{H}\rangle \left( 1+ \langle \kappa^\mathrm{LLS}\rangle / \langle \kappa^\mathrm{IGM} \rangle \right)$}, where $\langle \kappa^\mathrm{LLS}\rangle$ is the opacity due to LLS \citep{Prochaska2009} and $\langle \kappa^\mathrm{IGM}\rangle$ is the mean IGM opacity which is proportional to the mass weighted neutral fraction. As the Universe becomes more and more ionized,  $\langle \kappa^\mathrm{IGM}\rangle$ decreases and the contribution from $\langle \kappa^\mathrm{LLS}\rangle$ starts to become important. The dotted curves in the right panel of Figure~\ref{fig:fesc} show the $f_\mathrm{esc}^\mathrm{eff}$ obtained by using this modified equation. Only the estimate for \thesanlow simulation shows any real difference. This is because most other simulations have a late reionization model, where a part of the simulation volume is still neutral below $z=6$. Therefore, most of the opacity at these low redshifts is still provided by the IGM in an average sense. The impact of LLS opacity will only be large in the regions where the IGM is almost fully ionized. Therefore, the model fails to compensate for the LLS opacity in these `late' reionization models. 

\subsection{Bubble size statistics}

The statistical properties of ionized bubbles and their evolution with redshift will help place stringent constraints on the reionization process and the sources responsible for it \citep{Mellema2006}. This is because their size distribution depends on a range of physical processes that take place inside and around galaxies such as the radiation source function, feedback, recombinations, escape fractions etc. Observations of the 21\,cm line (from instruments like \lofar, \hera and \ska) will allow us to eventually get a full tomographic picture of the ionization field. This will for the first time help constrain the ionized bubble sizes in a fully statistical manner \citep{Eastwood2019}.

There are many different ways to quantify the bubble size statistics from a theoretical point of view, like the Friends of Friends algorithm \citep{Iliev2006}, spherical averaging schemes \citep{McQuinn2007}, mean free path method \citep{Mesinger2007}, watershed algorithms \citep{Lin2016}, granulometry \citep{grn, Busch2020} and more recently Minkowski tensors \citep{Kapahtia2021}. In this work we aim to quantify the bubble size distribution using the shot noise subtracted power spectrum of both ionized hydrogen (left panel of Figure~\ref{fig:Pk}) and the fluctuations in the brightness temperature of the redshifted 21cm signal (right panel of Figure~\ref{fig:Pk}). The value of the brightness temperature is given by \citep{Furlanetto2006, Choudhury2009}
\begin{equation}
\begin{split}
    T_b \simeq 22  \, x_\ion{H}{I} (1+\delta) \sqrt{\frac{1+z}{7}}
    \ \left( \frac{T_s - T_\gamma}{T_s}\right) \left[ \frac{H(z)/(1+z)}{\mathrm{d}v_{||}/\mathrm{d}r_{||}}\right] \, \mathrm{mK} \, ,
\end{split}
\end{equation}
where $x_\ion{H}{I}$ is the neutral fraction, $\delta$ is the gas overdensity, $T_s$ is the spin temperature, $T_\gamma$ is the CMB temperature and $\mathrm{d}v_{||}/\mathrm{d}r_{||}$ is the proper velocity gradient along the line of sight. In all our calculations we assume that the spin temperature is much greater than the CMB temperature \citep[$T_s \gg T_\gamma$; ][]{Pritchard2012} and that the \lyalpha coupling is sufficiently complete throughout the IGM \citep{WF2020}. This is a good approximation in the redshift range considered here \citep{Kulkarni2017}. 

 \begin{figure*}
	\includegraphics[width=0.495\textwidth]{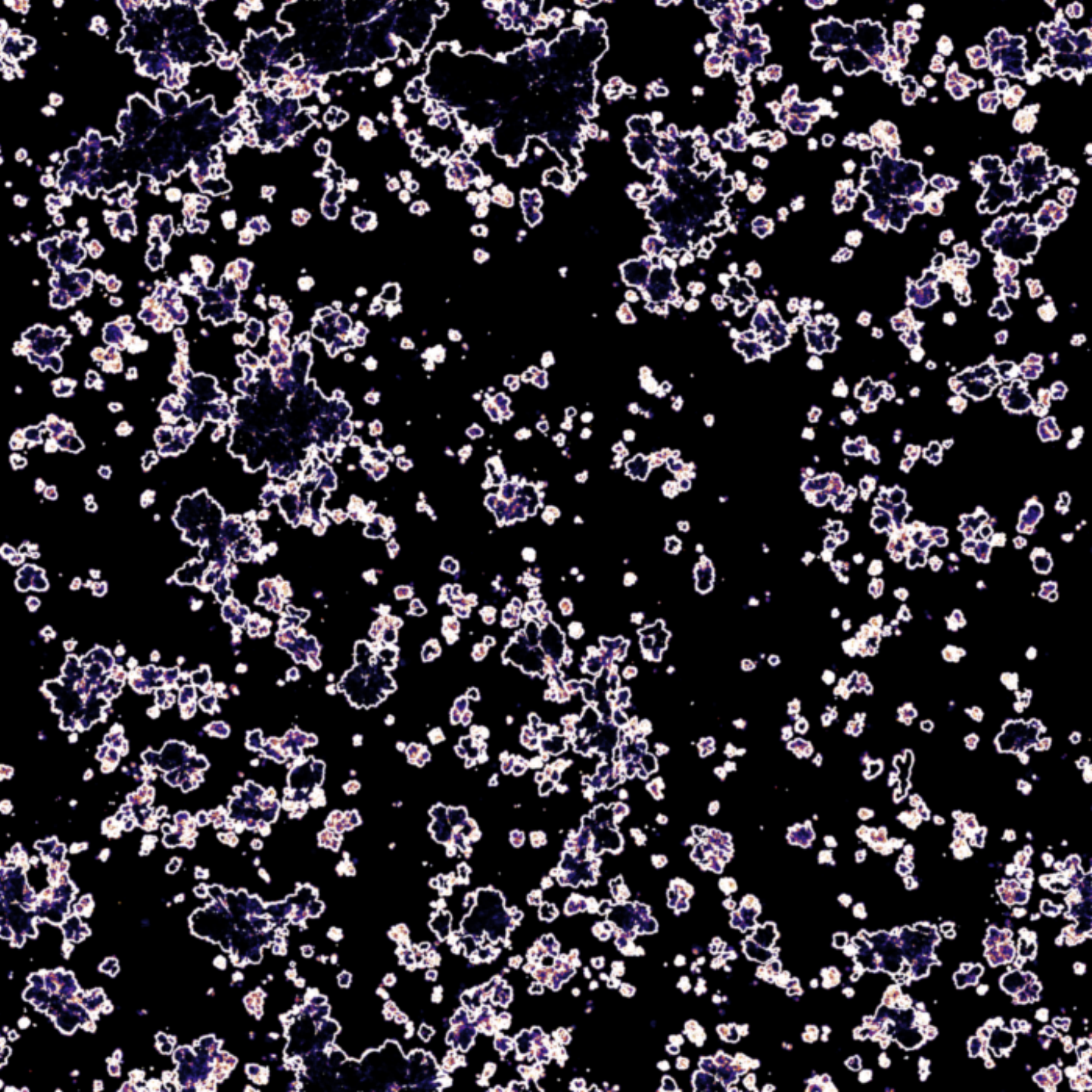}
	\includegraphics[width=0.495\textwidth]{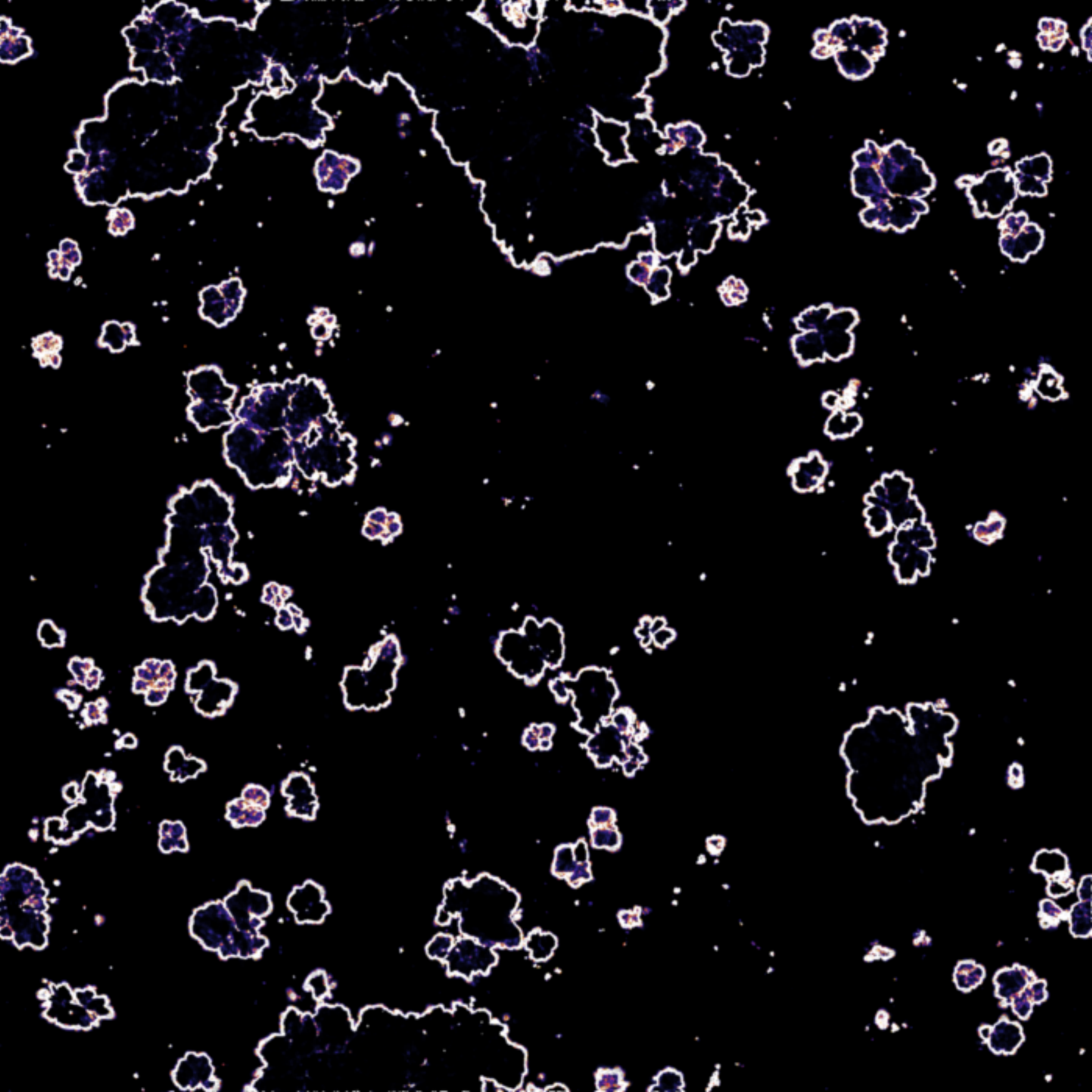}
    \caption{Bubble size distribution at a volume weighted ionization fraction of $0.3$ in the \thesanone (left panel) and \thesanhigh (right panel) simulations. The maps show the borders of the ionized bubbles obtained by plotting the pixels with the largest ionization gradients. It is clear that that similar values of ionization can be achieved in very different ways, with \thesanone exhibiting a large number of small bubbles and \thesanhigh containing less numerous but large \hii regions.}
    \label{fig:bubble_size}
\end{figure*}

 \begin{figure*}
	\includegraphics[width=0.999\textwidth]{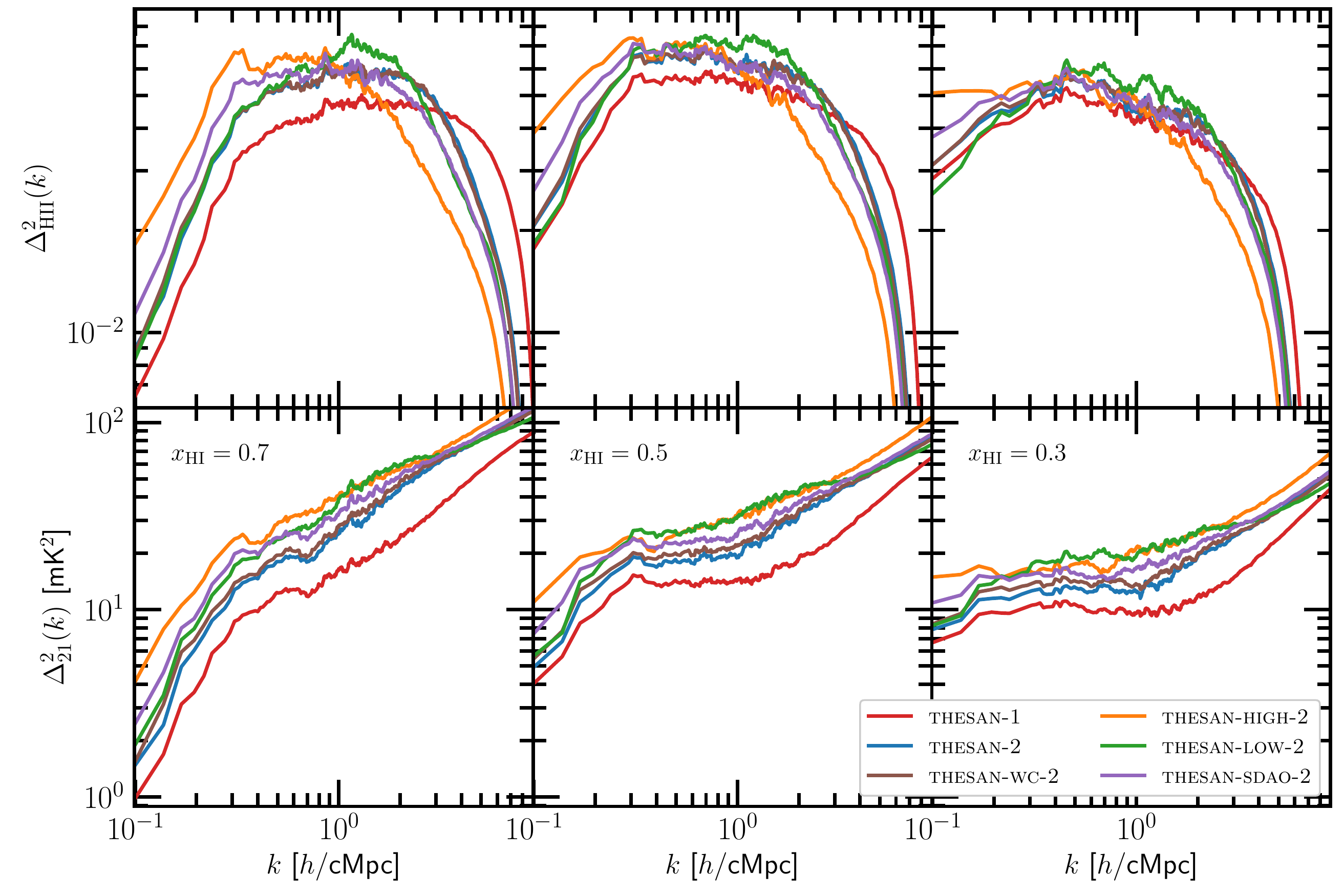}
    \caption{The power spectrum of ionized hydrogen (top panels)  and 21\,cm brightness temperature (bottom panels) for \thesanone (red curves), \thesantwo (blue curves), \thesanhigh (orange curves), \thesanlow (green curves) and \thesansdao (purple curves) at neutral fractions of $0.7$ (left panels), $0.5$ (middle panels) and $0.3$ (right panels). The different reionization models imprint uniquely distinct signatures on the ionized hydrogen and 21\,cm power spectra.}
    \label{fig:Pk_compare}
\end{figure*}

We first concentrate on the power spectrum (PS) of ionized hydrogen as this is a clean measurement and is fairly independent of the background matter density evolution. The \hii regions formed during the initial phases of the reionization process are quite small because the early sources of radiation were not very luminous. This is reflected in the PS, which peaks at large wavenumbers of about \mbox{$\sim 5$ $h$/cMpc} at $x_\ion{H}{II} =0.1$. As reionization progresses the ionized regions begin to become larger as the galaxies become bigger and the star formation rates increases. This shifts the peak to smaller and smaller wavenumbers. At the midpoint of reionization, the PS shows a  shallow plateau in the range \mbox{$k\simeq 0.3$--$3$ $h$/cMpc} implying that the bubbles encompass a wide range of sizes. Eventually, multiple nearby \hii regions overlap, shifting the power to the largest simulated scales. This is accompanied by a drop in the overall magnitude reflecting the drop in the fluctuations in the ionization field at the tail end of reionization.

The behaviour of the 21\,cm PS is a bit more complicated because the topology of the brightness temperature is a combination of the ionization field and the gas density distribution which depends on the cosmological parameters \citep{McQuinn2006, Kulkarni2017}. The evolution of the magnitude of the brightness temperature also complicates comparison between the PS derived at different redshifts. The small-scale power in the 21\,cm spectrum is largely dominated by the matter power spectrum. At very low ionization fractions ($<0.3$) even the large scale power appears to be dominated by the fluctuations in the matter density field. However, as the ionization bubbles become bigger, they start contributing power at large scales, creating a bump below \mbox{$k\lesssim 1$ $h$/cMpc}.   As ionized regions grow and the bubble sizes become larger the bump moves to smaller wavenumbers. Eventually almost all the power shifts to the largest spatial scale and the amplitude correspondingly decreases as the fluctuations in the ionization field decrease. Therefore, the position and amplitude of the bump can provide invaluable information about the reionization process \citep{Furlanetto2006}. The plot also shows the sensitivity limits for the current and upcoming second generation 21\,cm experiments like \lofar (blue lines), \hera (orange lines) and \ska (red lines) at $z=7$ (solid lines) and $z=8$ (dashed lines) \citep{Pober2014, Kulkarni2016}.  We note that \ska has the highest sensitivity but all three instruments should be able to measure the bump in the 21\,cm signal imprinted on it  by the reionization process. 

A major hurdle in the detection
of the 21\,cm signal is the accurate modelling and removal of foregrounds. For example, diffuse Galactic synchrotron radiation, supernova remnants, and extragalactic radio sources may be up to six orders of magnitude brighter than the targeted signal \citep{Dillon2014, Eastwood2019}. Fortunately, most of these sources are spectrally smooth allowing us to differentiate between the foregrounds and the underlying signal. Recent efforts from the first generations of instruments have placed upper limits on the signal that are about $2$--$4$ orders of magnitude larger than the values predicted by the \thesanone simulation \citep{Dillon2014, Ali2015, Jacob2015, Eastwood2019}. However, the increased sensitivity of
upcoming
instruments combined with improved foreground avoidance and subtraction schemes are expected to detect a definite 21\,cm signal during EoR in the near future \citep{HERA}.

\citet{Iliev2014} showed that, while simulations with box sizes of about \mbox{$\sim 100$ cMpc} are able to achieve convergent reionization histories, larger boxes (\mbox{$\gtrsim 200$ cMpc}) are necessary to obtain converged 21\,cm PS. However, recent higher resolution simulations \citep{Gnedin2014b, Kaurov2016} have shown that the bubble sizes and PS converge even for box sizes as low as \mbox{$\sim 60$ cMpc}. This discrepancy can potentially be explained by the fact that higher resolution allows for a reduction in the mean-free-path due to the opacity provided by more resolved LLS. This was not possible in the simulations presented in \citet{Iliev2014} where the resolution of the grid on which the RT equations were solved was about \mbox{$\sim 300$ ckpc}. Lower mean-free-paths limit the sizes of the largest bubbles, improving convergence even for smaller box sizes. Of course, we cannot rule out the presence or impact of 
bias due to the limited box size of our simulations. We hope to understand this behaviour better using upcoming larger volume simulations.

Up to this point we have focused on the evolution of the bubble sizes in \thesanone. It will, however, be very interesting to see how the bubble size distribution varies across the different simulations considered in this work. More importantly, if these differences manifest themselves in the PS, then upcoming observations might be able to place constraints on the different reionization models. Figure~\ref{fig:bubble_size} shows the outlines of ionized bubbles in a thin \mbox{($10$ cMpc)} slice in the \thesanone (left panel) and \thesanhigh (right panel) simulations at a volume weighted ionized fraction of $0.3$. It is immediately clear that similar values of ionization can be achieved in very different ways. \thesanone shows a large number of very small bubbles, while the majority of the ionized gas in \thesanhigh resides in one single large \hii region. At these low ionization fractions (high redshifts), the star fromation rate in \thesanone is dominated by very low mass halos (\mbox{$\Mh < 10^9 \Msun $}; Figure~\ref{fig:sfrd}). These halos, although numerous, are not very luminous and are less clustered \citep{Cooray2002}. Therefore, the \hii regions are small and well spread out throughout the entire simulation volume. On the other hand, the high mass halos are more clustered, as they only form in the highest density peaks, and they host galaxies that have very high star formation rates. Therefore, the radiation output from a single individual halo is very high.  This causes the bubble sizes to be large and less numerous than a model where the low mass halos drive the reionization process \citep{Lidz2008, Seiler2019}. 

Figure~\ref{fig:Pk_compare} quantifies the differences in bubble sizes between \thesanone (red curves), \thesantwo (blue curves), \thesanwc (brown curves), \thesanhigh (orange curves), \thesanlow (green curves) and \thesansdao (purple curves) simulations. It plots the PS of ionized hydrogen (top panels) and the 21\,cm emission (bottom panels)  at neutral fractions of $0.7$ (left panels), $0.5$ (middle panels) and $0.3$ (right panels). Encouragingly, there are very clear differences between the models considered. At $x_\ion{H}{I}=0.3$, the ionized hydrogen PS in \thesanone shows a wide array of bubble sizes in the range \mbox{$k\sim 0.3-6$ h/cMpc} with a peak at about \mbox{$k\sim 1 $ $h$/cMpc}. At the opposite end, \thesanhigh has primarily large bubbles peaking at \mbox{$k\sim 0.3$ $h$/cMpc} and very little power beyond \mbox{$k\sim 1 $ $h$/cMpc}. The lack of low mass halos in \thesansdao forces the higher mass halos to contribute more and produces relatively large bubbles which is reflected in the PS peaking at \mbox{$k\sim 0.5$ h/cMpc}. Owing to the fact that only low mass halos contribute, the PS in \thesanlow peaks at \mbox{$k\sim 1.5$ h/cMpc} and the distribution is quite narrow. Finally, \thesantwo  and \thesanwc show similar behaviours to \thesanone, except for the fact that they lack power at small spatial scales due to resolution effects suppressing star formation in low mass halos. As reionization progresses, all simulations show a consistent shift in power to smaller and smaller wavenumbers. This reflects the growth in bubble sizes due to increased star formation rates and merging. However, the differences between the models still remain with \thesanhigh and \thesansdao showing the highest power at small $k$ at any given ionization fraction. The picture at high-$k$ is a bit more complicated, but generally, the higher the contribution from low mass halos to reionization, the higher the wavenumber at which the PS falls of drastically.  

 \begin{figure}
	\includegraphics[width=0.99\columnwidth]{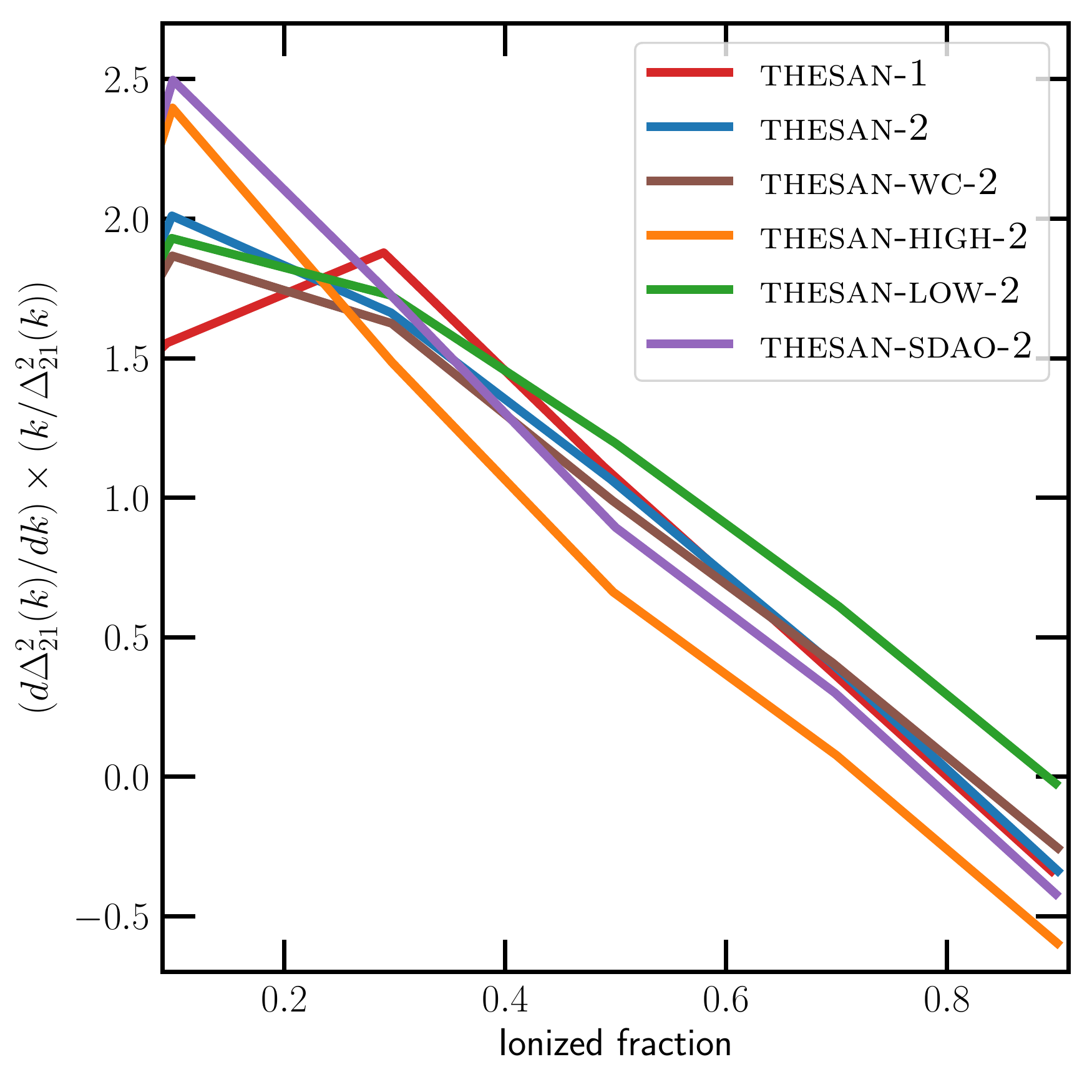}
    \caption{Dimensionless slope of the redshifted $21$ cm power spectrum at \mbox{$k=0.2$ h/Mpc} as a function of the volume weighted ionized fraction in the different simulations as indicated. Different reionization models exhibit noticeably different large scale slopes, with the simulations where the highest mass halos contribute the most number of ionizing photons to the reionization process showing the shallowest slopes and vice versa.}
    \label{fig:slope}
\end{figure}

This behaviour is also reflected in the 21\,cm PS. However, the picture is a bit more complicated because of the fact that the simulations reach different ionization fractions at different redshifts, therefore, the disparate matter density evolutions are also imprinted on this measurement. However, we can gain some insights by looking at just the shape (and disregarding the amplitudes) of the PS at scales where the power is not dominated by the matter power spectrum (\mbox{$k\lesssim 1$ $h$/cMpc}). At low ionization fractions, the bump caused by the reionization process is pretty small. However, the small-$k$ slope still encodes the information of \thesanhigh and \thesansdao having the largest bubble sizes. As ionized regions grow and the bubble sizes become larger the bump moves to smaller wavenumbers. This causes the small-$k$ slope to become shallower with the exact value and the rate at which it changes depending on the reionization properties of the simulation.

Figure~\ref{fig:slope} aims to quantify this behaviour by plotting the dimensionless slope of the PS as a function of the volume weighted ionized fraction at a wavenumber of \mbox{$k=0.2$ $h$/cMpc}. The amplitude and the evolution of the slope show very interesting differences between the different simulations. At a very low ionization fraction ($0.1$), the simulations in which low mass halos dominate the initial phase of the reionization process (\thesanone, \thesantwo, \thesanwc and \thesanlow),  have very small bubble sizes. Therefore, it does not affect (or very minimally affects) the PS on these large spatial scales causing the slope to be dominated by the matter power spectrum. \thesanhigh and \thesansdao simulations on the other hand start with relatively large bubble sizes and therefore the slope is already dominated by the ionization process with both of them showing a very shallow slope as expected. By \mbox{ $x_\ion{H}{II}=0.3$} all simulations are able to affect the slope of the PS on these scales. At this point, \thesanhigh shows the shallowest slope and \thesanone the steepest with the other simulations showing a slope in-between these two values. This is because, as discussed earlier, halos at the very low mass end contribute the most during the initial stages of reionization in \thesanone, while \thesantwo, \thesanwc and \thesanlow are unable to resolve these low mass halos. As reionization progresses, the slope in all simulations becomes shallower, eventually getting negative at high ionized fractions. \thesanlow has the steepest slopes above $x_\ion{H}{I}\sim0.4$, because the bubble sizes are consistently smaller than those in the other simulations. \thesanone,  \thesantwo and \thesanwc show (similar) slopes in-between the extreme values, because galaxies with a wide range of masses contribute to the reionization process. In general, the simulations where the high mass galaxies dominate (\thesanhigh; \thesansdao) the reionization process show consistently shallower slopes and vice versa. We note that instruments like \ska , can in the future, measure the slopes at these wavenumbers with enough sensitivity to distinguish between the different reionization models \citep{Seiler2019}.  It is therefore quite clear that the slope of the PS at small-$k$ and its evolution provides important information that can help place stringent constraints on the bubble size distribution, which in turn informs us about the sources that dominate the ionizing photon budget at any particular redshift.

\section{Conclusions}
\label{sec:conclusions}

We have introduced the \thesan project, a suite of large volume ($L_\mathrm{box} = 95.5 \, \mathrm{cMpc}$) radiation hydrodynamic simulations that self-consistently model the hydrogen reionization process and the sources responsible for it. The highest resolution simulation (\thesanone) has a DM mass resolution of $3.12 \times 10^6 \, \Msun$ and a baryonic mass resolution of $5.82 \times 10^6 \, \Msun$. The gravitational forces are softened on scales of $2.2~\ckpc$ with the smallest cell radii reaching $10$~pc. This allows us to model atomic cooling halos throughout the entire simulation volume.  The simulation uses well-tested galaxy formation (IllustrisTNG) and dust models to accurately predict the properties of galaxies that drive the reionization process. This is coupled to an efficient radiation hydrodynamics and non-equilibrium thermochemistry solver (\areport) that precisely captures the interaction between the radiation from galaxies and the surrounding low density IGM. Additionally, we also present a suite of lower resolution ($8$ times lower mass and $2$ times lower spatial resolution) simulations that help us to investigate the changes to reionization induced by halo mass dependent escape fractions (\thesanhigh and \thesanlow), alternative dark matter models (\thesansdao), assuming an instantaneous reionization model (\thesantng and \thesantngsdao), back reaction of the reionization process on galaxy formation (\thesannort), and numerical convergence (\thesantwo and \thesanwc). In this first paper, we have presented initial results focusing on reionzation histories, galaxy properties, escape fractions and bubble size statistics which are summarised below:
\begin{enumerate}
    \item The fiducial simulation and model variations all produce realistic reionization histories and match the observed neutral fraction evolution, the optical depth to CMB and the temperature of the IGM at mean density. \thesanone, \thesantwo and \thesanhigh have relatively large neutral regions below $z<6$, mimicking `late' reionization models, which have recently been invoked to explain long troughs in the \lyalpha forest. The duration of reionization is short in simulations where high mass halos  are the primary drivers of the reionization process (\thesanhigh and \thesansdao), while the opposite is true in simulations like \thesanone where low mass halos have a significant contribution. 
    \item On average, gas at higher overdensities reionizes earlier than low density gas in voids, supporting an inside-out reionization scenario where the ionization fronts that originate from galaxies slowly progress through the surrounding high-density gas and sweep through low-density regions at very high speeds. 
    \item The \thesan simulations are consistent with the estimated stellar to halo mass relation (SHMR) from \citet{Behroozi2019} and slightly below the estimates from \citet{Stefanon2021} but still consistent within the quoted errorbars. They also match the observed galaxy stellar mass function (SMF) at $z=6$--$10$. Comparable simulation efforts that model structure formation in the reionization epoch seem to have baryon conversion efficiencies that are either higher \citep[see for example, ][]{Rosdahl2018, Garel2021} or lower  \citep{Gnedin2014, Zhu2020} than the observationally inferred values.
    \item The dust-attenuated (using the empirical DTM relation derived in \citealt{Vogelsberger2020}) UV magnitudes of galaxies in the simulations match the observed luminosity function over a wide range. The observational uncertainties in the UVLF beyond \mbox{$M_\mathrm{UV}\sim -15$} are well represented in the simulations with the medium resolution runs showing a turnover at around the same magnitude as determined by \citet{Atek2018}. This is not present in \thesanone due to its superior resolution and is in better agreement with observations from \citet{Bouwens2017}.
    \item The star formation rate densities (SFRD) below $z\sim8$ are roughly converged between simulations and generally match the observationally inferred estimates. At higher redshifts, however, the lowest mass halos \mbox{($10^8\,\Msun < \Mh < 10^9\,\Msun$)}  dominate the total SFRD in \thesanone. This population is unresolved in the medium resolution simulations, reducing the SFRD by almost an order of magnitude. 
    \item The simulation volume is not large enough to get a statistically significant sample of high luminosity quasars. Moreover, the black-hole contribution to the total ionizing photon budget in \thesan is minimal ($<1\%$).
    \item The metal enrichment routines give rise to an accurate stellar mass-gas phase metallicity relation and the empirical dust model is in broad agreement with semi-analytic estimates at low metallicities. However, they predict relatively low dust-to-metal and dust-to-gas ratios at high metallicities. The metals and dust produced in the ISM are efficiently transported out by feedback driven winds, thus significantly enriching both the CGM and IGM by $z=6$. 
    \item The computed emissivities and clumping factors are in broad agreement with previous simulations. Using a simple analytic model \citep[first described in ][]{Madau1999} we estimate that the effective escape fraction of ionizing photons is about $15-20\%$ in our fiducial model, which is, again, in agreement with previous estimates. Although these calculations require quite a few approximations, like assuming a Case B recombination rate, a constant ionized gas temperature of \mbox{$10^4$ K} and that $\mathcal{C}_{100}$ is representative of the amount of clumping in the gas that is recombining, it does provide important predictions for the analytic reionization models widely used in the literature \citep[for e.g.][]{Madau2017}. A thorough investigation of more precise luminosity-weighted escape fractions using ray-tracing codes will be performed in a forthcoming paper.
    \item The ionization process contributes power at large scales to the 21\,cm spectrum, creating a bump below \mbox{$k\lesssim 1$ $h$/cMpc}. As ionized regions grow and the bubble sizes become larger the bump moves to smaller wavenumbers. Eventually almost all the power shifts to the largest spatial scales and the amplitude decreases as the fluctuations in the ionization field subside.  Importantly, upcoming instruments like \lofar, \hera and \ska have enough sensitivity to detect these reionization signatures in the 21\,cm power spectrum. 
    \item Different reionization models produce different bubble size distributions at the same ionization fraction. Simulations where low mass halos dominate the ionizing output form a large number of very small bubbles, while simulations in which the high mass halos contribute the dominating number of ionizing photons form less numerous but large \hii regions. Encouragingly, these different bubble sizes show unique signatures in the 21\,cm PS. We show that the slope at small-$k$ \mbox{($k=0.2$ $h$/cMpc)} is able to differentiate between the different reionization models and  can, therefore, help to place stringent constraints on the bubble size distribution and evolution, which in turn informs us about the sources that dominate the ionizing photon budget.
\end{enumerate}

We note that while the \thesan simulations use the same underlying galaxy formation model as IllustrisTNG, some important differences exist. For example, we replace the spatially uniform UV background with full RT, the equilibrium thermochemistry solver is replaced with a non-equilibrium one and we also include an empirical dust model. These changes can in principle alter gas cooling rates and the metal content in galaxies. Additionally, the differences in the mass and spatial resolution can also lead to differences in predicted galaxy properties, because the galaxy formation model is not perfectly converged with resolution. For example, a higher resolution generally leads to slightly higher baryonic conversion efficiencies resulting in larger stellar masses for a fixed halo mass \citep{Pillepich2018}. Although we have checked that the properties of galaxies at high-$z$ ($z\gtrsim5.5$) are not significantly affected, the impact of these changes at lower redshifts is unclear, and it is possible that there might be some noticeable differences between the low-$z$ galaxy populations predicted by IllustrisTNG and \thesan.

We have shown that large scale coupled radiation hydrodynamic and galaxy formation simulations have the ability to model both the large scale statistical properties of the IGM during the reionization process and the resolved characteristics of the galaxies responsible for it. They are, therefore, able to make predictions for, and help interpret observations from, a slew of current and future telescopes such as the \jwst, \alma, \hera, \ska, \texttt{CCAT-p}, \texttt{SPHEREx}, etc., that have been specifically designed to study high-redshift structure formation and EoR. In accompanying papers, we study the IGM using the \lyalpha forest, quantify \lyalpha IGM transmission, make predictions for nebular line intensity mapping experiments, explore escape fraction statistics, quantify the back reaction of reionization on galaxy formation, and so forth. We therefore hope to significantly advance our understanding of high-redshift structure formation and EoR using the \thesan simulations.

\section*{Acknowledgements}
We thank the referee Harley Katz for constructive comments that helped improve the quality of this paper. We thank Ewald Puchwein, Sandro Tacchella, Sebastian Bohr and Benedetta Ciardi for useful comments and discussions. AS acknowledges support for Program number \textit{HST}-HF2-51421.001-A provided by NASA through a grant from the Space Telescope Science Institute, which is operated by the Association of Universities for Research in Astronomy, incorporated, under NASA contract NAS5-26555. MV acknowledges support through NASA ATP grants 16-ATP16-0167, 19-ATP19-0019, 19-ATP19-0020, 19-ATP19-0167, and NSF grants AST-1814053, AST-1814259,  AST-1909831 and AST-2007355. The authors gratefully acknowledge the Gauss Centre for Supercomputing e.V. (\url{www.gauss-centre.eu}) for funding this project by providing computing time on the GCS Supercomputer SuperMUC-NG at Leibniz Supercomputing Centre (\url{www.lrz.de}). Additional computing resources were provided by the Extreme Science and Engineering Discovery Environment (XSEDE), at Stampede2 through allocation TG-AST200007  and by the NASA High-End Computing (HEC) Program through the NASA Advanced Supercomputing (NAS) Division at Ames Research Center. We are thankful to the community developing and maintaining software packages extensively used in our work, namely: \texttt{matplotlib} \citep{matplotlib}, \texttt{numpy} \citep{numpy} and \texttt{scipy} \citep{scipy}.

\section*{Data Availability}

 All simulation data, including snapshots, group and subhalo catalogues, merger trees, and high time cadence Cartesian outputs will be made publicly available in the near future. Data will be distributed via \url{www.thesan-project.com}. Before the public data release, data underlying this article will be shared on reasonable request to the corresponding author(s).


\bibliographystyle{mnras}
\bibliography{bibliography} 



\appendix

\section{Effect of reduced speed of light approximation}

\label{app:RSLA}
 \begin{figure}
	\includegraphics[width=0.99\columnwidth]{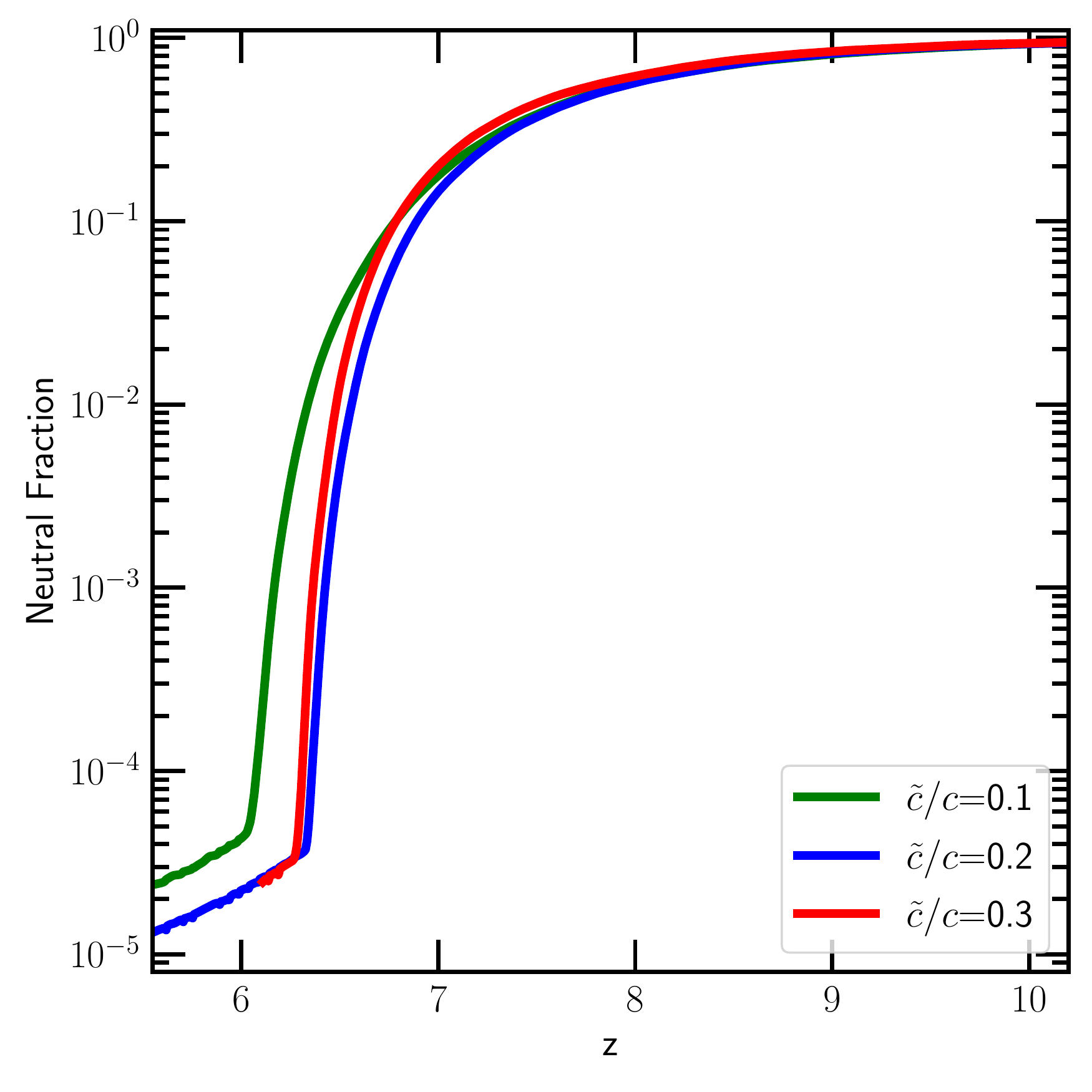}
    \caption{Effect of a reduced speed of light. Convergence on the evolution of the neutral fraction is achieved with a value of $\tilde{c} = 0.2 c$.  }
    \label{fig:c_reduced}
\end{figure}

In fully coupled radiation hydrodynamic simulations the most computationally intensive part of the calculation arises from the small timesteps required to accurately model the high signal speed of light. Therefore, most simulations use a reduced speed of light approximation, which works as long as the speed of the ionization fronts are slower than the reduced light speed \citep{Gnedin2001}. In general, in galaxy formation simulations, which mainly focus on high density regions in and around halos, light speeds as low as \mbox{$1000\,\text{km\,s}^{-1}$} are sufficient to achieve converged results \citep{Kannan2020a, Kannan2020b}. However, at the tail end of reionization, the I-fronts can reach speeds as high as $0.1 c$ as they sweep through the low density voids \citep{Daloisio2019}. Recent investigations have shown that some semblance of convergence can be achieved (at least for relatively low simulation volumes) by assuming $\tilde{c} = 0.3c$ \citep{Deparis2019} although this will still not get accurate post reionization neutral gas fractions (\citealt{Ocvirk2019}; however see \citealt{Gnedin2016} for an alternative implementation of the reduced speed of light approximation that recovers accurate post-reionization neutral fractions even at low speeds of light). Recent works that model reionization using \areport have shown that the timing of reionization is converged if a value of $\tilde{c} = 0.3c$ is used \citep{Wu2019a} and the IGM heating is converged even with a value as low as $0.1c$ \citep{Wu2019b}.

We show a similar test in Figure~\ref{fig:c_reduced}, which depicts the ionization fraction evolution in a $50$ \cMpc\xspace volume using the fiducial model but at a lower resolution containing $2\times512^3$ particles. The simulation with $\tilde{c} = 0.1c$ shows a delay in the reionization process particularly caused by slow I-front speeds during the end stages of reionization. In contrast, the simulations with $\tilde{c} = 0.2c$  and $\tilde{c} = 0.3c$ seem well converged. We therefore use a value of $\tilde{c} = 0.2c$ in all the \thesan simulations and note that this value has also been used in other recent works that model the reionization process \citep{Rosdahl2018}. 

\bsp	
\label{lastpage}
\end{document}